\documentclass[fleqn,twoside]{article}
\usepackage{espcrc2}

\usepackage{graphicx}
\usepackage[figuresright]{rotating}
\usepackage{amsmath}
\usepackage{amssymb}
\usepackage{mathrsfs}

\def\slc#1{\setbox0=\hbox{$#1$}           % set a box for #1
    \dimen0=\wd0                                 % and get its size
    \setbox1=\hbox{/} \dimen1=\wd1               % get size of /
    \ifdim\dimen0>\dimen1                        % #1 is bigger
       \rlap{\hbox to \dimen0{\hfil/\hfil}}      % so center / in box
       #1                                        % and print #1
    \else                                        % / is bigger
       \rlap{\hbox to \dimen1{\hfil$#1$\hfil}}   % so center #1
       /                                         % and print /
    \fi}

\newcommand{\bi}{\begin{itemize}}
\newcommand{\ei}{\end{itemize}}

\newcommand{\be}{\begin{equation}}
\newcommand{\ee}{\end{equation}}
\newcommand{\bea}{\begin{eqnarray}}
\newcommand{\eea}{\end{eqnarray}}

\newcommand{\ldm}{\Delta m_{31}^2}
\newcommand{\sdm}{\Delta m_{21}^2}
\newcommand{\deltacp}{\delta_{\mathrm{CP}}}
\newcommand{\stheta}{\sin^2 2 \theta_{13}}

\newcommand{\ie}{{\it i.e.}}

\newcommand{\eg}{{\it e.g.}}

\newcommand{\cf}{{\it cf.}}

\newcommand{\eq}{Eq.}

\newcommand{\fig}{Fig.}

\newcommand{\Ref}{Ref.}
\newcommand{\Refs}{Refs.}
\newcommand{\Sec}{Sec.}

\newcommand{\Tab}{Table}

\newcommand{\equ}[1]{\eq~(\ref{equ:#1})}
\newcommand{\figu}[1]{\fig~\ref{fig:#1}}

\newcommand{\hl}[1]{{\em #1}}

\newcommand{\GLOBES}{{\rm GLoBES}}
\newcommand{\AEDL}{{\rm AEDL}}

\title{Lectures on neutrino phenomenology}

\author{Walter Winter\address[Wue]{Institut f{\"u}r theoretische Physik und Astrophysik, 
        Universit{\"a}t W{\"u}rzburg, \\ Am Hubland, D-97074 W{\"u}rzburg, Germany}%
        \thanks{This work has been supported by the Emmy Noether program of Deutsche Forschungsgemeinschaft.}}
       
\begin{document}

\begin{abstract}

The fundamental properties of the lepton sector include the neutrino masses and flavor mixings. Both are difficult to observe because of the extremely small neutrino masses and neutrino-matter cross sections. 
In these lectures, we focus on the basic concepts for the determination of neutrino properties. We introduce neutrino oscillations as standard mechanism for neutrino flavor changes, and we discuss methods to measure the neutrino mass. Furthermore, we illustrate how precision measurements in neutrino oscillations will be performed in the future, and may even open a window to new physics properties, such as motivated by LHC physics. Finally, we discuss some applications of neutrinos in astrophysics, such as neutrino oscillations in the Sun. We also illustrate how neutrinos from extragalactic cosmic accelerators may be used for the determination of neutrino properties. 

\vspace{1pc}
\end{abstract}

\maketitle

\section{Introduction}

Neutrinos are the most abundant known matter particles in the universe, in number, exceeding the  constituents of ordinary matter (electrons, protons, neutrons) by a factor of ten billion. The neutrino fluxes are extremely high. For example, there are about $7 \cdot 10^{10} \, \mathrm{cm}^{-2} \, \mathrm{s}^{-1}$ streaming through the Earth from the Sun.  Neutrinos are naturally produced in the Big Bang, the Earth's crust (by uranium and thorium decays), the Sun, supernovae, and the Earth's atmosphere by cosmic ray interactions. Furthermore, at the highest energy scales, they are most probably produced in cosmic accelerators together with the observed cosmic rays. Man-made neutrinos are emitted from nuclear fission reactors, as well as it is possible to produce artificial neutrino beams. Although they are very abundant, it is difficult to catch them because of their extremely small cross sections. As a consequence, they even escape from very dense environments, such as supernovae or nuclear power plants. Roughly speaking, the number of detected neutrinos scales as
\begin{equation}
N_{\mathrm{obs}} \propto \phi \times \sigma \times t \times m_{\mathrm{Det}} \,.  
\label{equ:sizedet}
\end{equation}
The flux $\phi$ is extremely large, the cross section $\sigma$ extremely small. The observation time should not be longer than a few years. In order to accumulate sufficient statistics, one therefore can estimate the necessary detector sizes from \equ{sizedet} to be of order of kilotons.
In the Standard Model (SM)  of elementary particle physics, neutrinos are massless particles. However,  after a long history of disputed alternatives, neutrino  oscillations have recently been established as leading neutrino flavor change mechanism, which implies that at least two of the neutrino (mass) states must be massive. Massive neutrinos are now one of the very few specific hints for physics beyond the SM. 

Although the SM can be easily extended by right-handed neutrinos to introduce Dirac mass terms, the lightness of the neutrinos then points to un-plausibly small Yukawa couplings. A typical way out is the introduction of a Majorana neutrino mass, which implies that the neutrino is its own antiparticle. This alternative has a number of interesting implications. First of all, this hypothesis can be, in principle, tested in neutrinoless double beta ($0\nu\beta\beta$) decay, unless it is suppressed by intricate parameter constellations. Second, a Majorana mass term violates lepton number, which is an accidental symmetry of the SM. Such a lepton (or baryon) number violation is, for instance, needed in dynamical mechanisms to describe the observed matter-antimatter-asymmetry in the universe (baryogenesis). Maybe neutrinos are somehow involved in such a mechanism (leptogenesis~\cite{Fukugita:1986hr}). And third, neutrino mass can be interpreted as the lowest order perturbation of Beyond the SM (BSM) physics in the sense of the effective operator picture.  In this picture,  the famous Weinberg-Operator~\cite{Weinberg:1979sa}
\begin{equation}
\mathscr{L}^{d=5}_{\text{eff}} \propto \frac{1}{\Lambda}  \,
\left(L \, i \sigma^2 H \right) \, \left(L \, i \sigma^2 H \right) + \text{h.c.}
\label{equ:weinberg}
\end{equation}
($L$: lepton doublet, $H$: Higgs doublet) is the only $d=5$ operator, and it is the only operator suppressed by only one power of $\Lambda$, where $\Lambda$ is the BSM physics scale. It leads to Majorana neutrino masses after electroweak symmetry breaking (EWSB). Using order one couplings and neutrino masses $\lesssim \mathrm{eV}$, one can easily show that $\Lambda$  in \equ{weinberg} points towards the Grand Unified Theory (GUT) scale, the scale where the gauge interactions of the SM are expected to unify. In this case, it may be obvious to suspect a connection with the mentioned baryogenesis concept in the early universe. The fundamental theories leading to the operator in \equ{weinberg} and their implications are discussed in the lecture by Z.-z.~Xing in this series~\cite{XingLecture} (see also \cite{FritschLecture} for some aspects). 

In order to describe possible theories for neutrino mass and the connected BSM physics, it is mandatory to pin down the properties of the neutrinos. As we shall see, most of the observables, such as the mass (squared) splittings and mixing parameters, are accessible in neutrino oscillations. In particular, the observation of leptonic CP violation may support the argument that neutrinos are involved in leptogenesis. Neutrino oscillations can be used to fix essentially all expected (guaranteed) observables except from the absolute neutrino mass scale and possible Majorana phases. The direct measurement of the neutrino mass typically involves the determination of the endpoint in the  electron spectrum coming from tritium decays. The direct test of the nature of neutrino mass involves the test of $0\nu\beta\beta$ decay. In addition, upper bounds on the neutrino mass scale can be obtained from cosmology. Apart from neutrino flavor mixing and neutrino masses, other properties of the neutrinos may point towards the nature of neutrino mass, such as the electromagnetic dipole moment, or to new physics effects, such as neutrino lifetime. Therefore, it is worthwhile to test and constrain these as well. Especially neutrino from extragalactic cosmic accelerators may allow for probes in different baseline and energy regimes, where new physics effects may be present which are otherwise not observable.
 
In these lectures titled ``neutrino phenomenology'', we focus on the determination of the neutrino properties. In \Sec~\ref{sec:so}, we discuss the current understanding of neutrino oscillations. Then in \Sec~\ref{sec:prec}, we show future perspectives for neutrino oscillation experiments, with special emphasis on the discovery of leptonic CP violation. In this section, we add matter effects in (constant) Earth matter to the framework of neutrino oscillations. As the next step, we illustrate how neutrino oscillations work in varying matter density in the Sun in \Sec~\ref{sec:sun}, and show how these can be used to test the solar neutrino mixing angle. In \Sec~\ref{sec:cosm} we discuss the possibility to use neutrinos from cosmic accelerators for the test of neutrino properties, such as neutrino lifetime. Finally, we summarize the approaches to test the absolute neutrino mass scale in \Sec~\ref{sec:mass}. 

\section{Neutrino oscillation framework}
\label{sec:so}

Here we present the current understanding of neutrino oscillations. We start with a short historical perspective, then introduce neutrino oscillations in vacuum in the standard quantum mechanical treatment, and comment on leptonic CP violation. Then we derive the two-flavor limit, and demonstrate that the general three-flavor case can be reduced to two-flavor sub-sectors for our current understanding. 

\subsection{Historical perspective}

Historically, neutrinos from the Sun were observed for many decades by Raymond Davis, Jr. and collaborators since the 1970s in the Home\-stake experiment~\cite{Cleveland:1998nv}. On the other hand, solar models based on the nuclear fusion chains in the Sun, such as by John N. Bahcall and collaborators (see, \eg, \Ref~\cite{Bahcall:2004pz} for a recent discussion), predicted much higher (electron neutrino) fluxes based on the normalization to the solar luminosity. This solar neutrino anomaly  (see \Ref~\cite{Bahcall:1976zz} for an early reference) can be plausibly described by neutrino flavor changes. The most important results in solar neutrino physics in this context is probably the SNO (Sudbury Neutrino Observatory) neutral current measurement~\cite{Ahmad:2002jz}, which confirmed the flux predictions of the solar model, and therefore the flavor changes of the neutrinos. Since the neutral current interactions are equally sensitive to all (active) neutrino flavors, the ratio between  charged current interactions of electron neutrinos and neutral current interactions directly determines the fraction of neutrinos still found in the original state. Apart from neutrinos from the Sun, neutrinos are abundantly produced in the Earth's atmosphere. Cosmic rays, which could be protons or heavier nuclei, hit the Earth's atmosphere to produce showers including charged pions. These charged pions decay with the decay chain
\begin{eqnarray}
\pi^+ & \rightarrow & \mu^+ + \nu_\mu  \, ,\nonumber \\
& &  \mu^+ \rightarrow e^+ + \bar\nu_\mu + \nu_e \, ,
\label{equ:pion}
\end{eqnarray}
as illustrated for $\pi^+$ here, leading to a flux of electron and muon neutrinos and antineutrinos. The most compelling evidence for neutrino oscillations as leading flavor change mechanism is probably coming from the observation of atmospheric neutrinos in the Super-Kamiokande experiment~\cite{Fukuda:1998mi}. This experiment has used the directional dependence of the incoming neutrinos, which determines the path length traveled through the Earth, to infer the neutrino oscillation parameters. In fact, it has turned out that solar and atmospheric neutrino flavor changes can be described by different sets of oscillation parameters. The corresponding oscillations can be described as sub-sectors of the general three-flavor case. 

In this section, we choose a deductive rather than historical presentation.  We show that the existing measurements can be described by leading two-flavor sub-sectors, which our understanding is based upon. Although such a presentation suggests  that neutrino oscillations have been taken for granted as flavor change mechanism, they have been established by considering a number of alternatives, such as neutrino decay and decoherence, and a number of anomalies. 
For a more refined presentation of the current picture, see \Ref~\cite{GonzalezGarcia:2007ib}, and for a more detailed presentation of the neutrino oscillation phenomenology, see, \eg, \Ref~\cite{Giunti:2007ry} and references therein. In the next section, we will then look beyond these two-flavor sectors and introduce  three-flavor effects. Although some of the recent experiments are already sensitive to three-flavor effects, we perform this splitting for conceptual reasons. 

\subsection{Neutrino oscillations in vacuum}

In a quantum mechanical picture, the eigenstates of the weak charged current interactions $| \nu_\alpha \rangle$ do not correspond to the mass eigenstates $| \nu_k \rangle$, \ie, the eigenstates of the free Hamiltonian $\mathcal{H}$ with the neutrino energy eigenvalues $E_k = \sqrt{\vec p^2 + m_k^2}$. The flavor and mass eigenstates are connected by a unitary $(N+S) \times (N+S)$ matrix 
\begin{equation}
| \nu_\alpha \rangle = \sum\limits_{k=1}^{N+S} U_{\alpha k}^* |\nu_k \rangle \, ,  \label{equ:mix}
\end{equation}
where $N$ is the number of active and $S$ is the number of sterile (\ie, not weakly interacting) neutrino mass eigenstates.\footnote{The contribution of sterile states is strongly constrained. However, we keep them in the derivation to demonstrate where the calculation works independently of the number of participating flavors.} Greek indices denote flavor eigenstates, Latin indices denote mass eigenstates. The unitary mixing matrix $U$ is, for $N=3$ and $S=0$, also called $U_{\text{PMNS}}$, or \hl{Pontecorvo-Maki-Nakagawa-Sakata} matrix. The flavor and mass eigenstates, respectively, are both assumed to form a basis.
Applying the time evolution of the mass eigenstates in vacuum
\begin{equation}
| \nu_k (t) \rangle = \exp (- i E_k t)  \, | \nu_k \rangle  
\end{equation}
to \equ{mix} and using the unitarity of the mixing matrix, we obtain the vacuum transition amplitude
\begin{eqnarray}
 A_{\nu_\alpha \rightarrow \nu_\beta} & \equiv & A_{\alpha \beta} =  \langle \nu_\beta | \nu_\alpha (t) \rangle  \\
& = & \sum\limits_{k=1}^{N+S} U_{\alpha k}^* U_{\beta k} \, \exp( - i E_k t ) \, . \nonumber
\label{equ:a}
\end{eqnarray}
Note that this equation can be equivalently obtained from the Schr{\"o}dinger equation in matrix form
\begin{equation}
i \frac{d}{dt} \Psi = \mathcal{H}_F \, \Psi \, , \quad \mathcal{H}_F= U \, 
\left( \begin{array}{ccc} E_1 & 0 & \hdots \\ 0 & E_2 &  \\ \vdots &  & \ddots \end{array} \right) U^\dagger \, 
\label{equ:ham}
\end{equation}
with $\mathcal{H}_F$ the Hamiltonian in flavor space and $\Psi$ the flavor state vector. This form is often used if the Hamiltonian is explicitely time-dependent, such as it might be for the matter effects discussed later.
For the transition probability, we have from \equ{a}
\begin{eqnarray}
P_{\alpha \beta} & = & A_{\alpha \beta}^* A_{\alpha \beta} =  \label{equ:oscpre} \\
& = & \sum\limits_{k, \, j = 1}^{N+S} \underbrace{U_{\alpha k}^* U_{\beta k} U_{\alpha j} U_{\beta j}^*}_{\equiv J_{kj}^{\alpha \beta}} \, e^{ - i (E_k - E_j) t} \, . \nonumber
\end{eqnarray}
The quantity $J_{kj}^{\alpha \beta}$ is also known as \hl{quartic re-phasing invariant}~\cite{Jenkins:2007ip}, which characterizes the information in the mixing matrix independent of a possible phase re-definition of the charged lepton and neutrino fields. The standard derivation of the oscillation formula relies on the approximations for ultra-relativistic neutrinos
\begin{equation}
E_k = \sqrt{\vec p^2 + m_k^2} \simeq E + \frac{m_k^2}{2 E} \, , \quad t \simeq L \, , 
\label{equ:ur}
\end{equation}
which imply that the different mass eigenstates have different energies. This seems to be contradictory to a neutrino produced with a certain energy as a superposition of mass eigenstates. Therefore, in any more refined calculation, assumptions on the energy and momentum widths have to be made. The simplest such method includes the assumption of wave packets, see, \eg, \Ref~\cite{Giunti:1997wq}, which leads to the same oscillation formula we will obtain -- as long as there is enough wave packet overlap among the different mass eigenstates.
From \equ{oscpre} using \equ{ur} and the definition $\Delta m_{kj}^2 \equiv m_k^2 - m_j^2$, we 
find after some transformations the neutrino oscillation probability in vacuum
\begin{eqnarray}
P_{\alpha \beta} & = &\delta_{\alpha \beta} - \underbrace{4 \, \sum\limits_{k > j} \mathrm{Re}  J_{kj}^{\alpha \beta} \, \sin^2 \left( \frac{\Delta m_{kj}^2 L}{4 \, E} \right)}_{\mathrm{CP \, conserving}} \nonumber \\
& + & \underbrace{2 \, \sum\limits_{k > j} \mathrm{Im}  J_{kj}^{\alpha \beta} \, \sin \left( \frac{\Delta m_{kj}^2 L}{2 \, E} \right)}_{\mathrm{CP \, violating}} \, . \label{equ:osc}
\end{eqnarray}
The quantity $L$, the distance between source and detector, is often called \hl{baseline}, and the functional dependence  $\mathcal{F}(L,E)$ on $L$ and $E$ is often called \hl{spectral dependence}; in vacuum, it is just $\mathcal{F}(L,E)=L/E$,
as one can read off from \equ{osc}. The evidence for neutrino oscillations as leading flavor change effect comes from this particular spectral dependence, \ie, the flavor change effect as a function of energy and baseline. 
There is no sensitivity of neutrino oscillations to the absolute mass scale, but the mass (squared) splittings and the ordering of the masses is, in principle, determined by \equ{osc}. For $N+S$ flavors, there are $N+S-1$ independent mass squared splittings.

\subsection{On leptonic CP violation}

A very important quantity of interest is the \hl{CP symmetry}, where ``CP'' stands for charge-parity. It is basically a symmetry between the behavior of particles and anti-particles taking into account the peculiarities of the electroweak framework (in particular, the $V-A$ interactions only coupling to left-handed particles and right-handed anti-particles). In the context of our CP asymmetric universe in which we do not find sufficient antimatter to justify the CP symmetry, the question of the source of the violation of this symmetry, in short, \hl{CP violation}, is probably one of the most interesting ones in particle physics.
The CP conserving part in \equ{osc} is the same for neutrinos and anti-neutrinos. The CP violating part changes sign for anti-neutrinos (for which the mixing matrix effectively has to be complex conjugated). It is only present if the mixing matrix has complex phases (apart from possible Majorana phases) invariant under a re-definition of the lepton fields. 
Note that the CP violating part oscillates with the double frequency compared to the CP conserving part. In addition, note that $\mathrm{Im}  J_{kj}^{\alpha \beta}$ is, up to a sign depending on the indices, equivalent to the so-called \hl{Jarlskog invariant} $\mathcal{J}$~\cite{Jarlskog:1985ht}, which is frequently used for the quantification of CP violation. Another quantity, which has been used historically, is the \hl{CP asymmetry}
\begin{equation}
a_{\mathrm{CP}} = \frac{P_{\alpha \beta}-P_{\bar\alpha \bar\beta}}{P_{\alpha \beta}+P_{\bar\alpha \bar\beta}} \,  ,
\label{equ:cpasym} 
\end{equation}
where $P_{\bar\alpha \bar\beta}$ refers to $P_{\bar\nu_\alpha \rightarrow \bar\nu_\beta}$. For modern statistical simulations, this quantity is not very representative, mainly because of matter effects, which violate the CP symmetry extrinsically by the absence of antimatter in the Earth, as we will discuss in the next section. However, one easily finds from \equ{osc} and $J_{kj}^{\alpha \alpha} = |U_{\alpha k}|^2 \, |U_{\alpha j}|^2$ that $a_{\mathrm{CP}}=0$ for $\alpha=\beta$, \ie, one needs to observe the transition among flavors to access CP violation.
If the oscillations are averaged out, such as by a poor energy resolution of the detector, we have 
for the oscillatory terms in \equ{osc} 
\begin{eqnarray}
\left< \sin^2 \left( \frac{\Delta m_{kj}^2 L}{4 E} \right) \right>_{L/E} = \frac{1}{2} \, , \nonumber \\
\left< \sin \left( \frac{\Delta m_{kj}^2 L}{2 E} \right) \right>_{L/E} = 0 \, .
\label{equ:average}
\end{eqnarray}
This implies that the measurement of CP violation requires in addition the observation of the spectral dependence.

\begin{figure}[t!]
   \centering \includegraphics[width=\columnwidth]{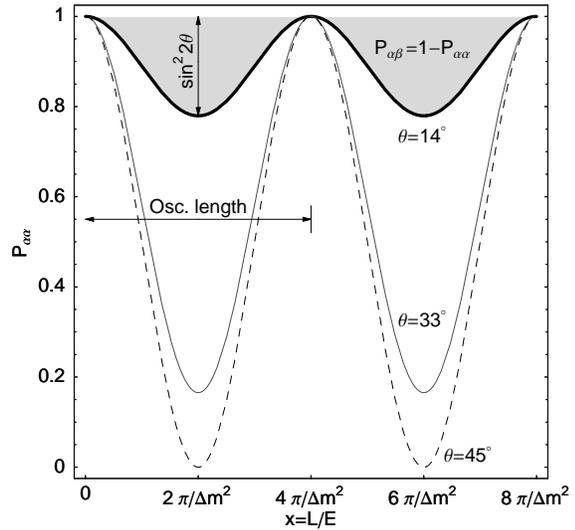}

   \vspace*{-0.7cm}

   \caption{\label{fig:oscprob} Disappearance probability $P_{\alpha \alpha}$ as a function of $x=L/E$ in the two-flavor limit. The different curves represent different values of the mixing angle for illustration. The shaded region corresponds to the appearance probability $P_{\alpha \beta}=1-P_{\alpha \alpha}$ (for $\theta=14^\circ$).}
\end{figure}

\subsection{Two-flavor limit}

In order to illustrate neutrino oscillations, it is useful to consider the \hl{two-flavor limit}, \ie, $N=2$ and $S=0$. From the simple two-flavor mixing matrix 
\begin{equation}
U = \left( \begin{array}{cc} \cos \theta & \sin \theta \\ -\sin \theta & \cos \theta \end{array} \right) \, ,
\label{equ:utwo}
\end{equation}
parameterized by only one mixing angle $\theta$, we directly obtain from \equ{osc} the transition probabilities for the two flavor states $| \nu_\alpha \rangle$
and $| \nu_\beta \rangle$ separated by only one  mass squared splitting $\Delta m^2$:
\begin{eqnarray}
P_{\alpha \alpha} & = & 1 - \sin^2  2 \theta  \,  \sin^2 \left( \frac{ \Delta m^2 L}{4 E} \right) \, , \nonumber \\
P_{\alpha \beta} & = & \sin^2  2 \theta  \,  \sin^2 \left( \frac{ \Delta m^2 L}{4 E} \right) \, .
\label{equ:twoflavor}
\end{eqnarray}
The first probability is often called \hl{disappearance probability} or \hl{survival probability} (because the flavor $\nu_\alpha$ disappears or survives, depending on the point of view), and
the second probability is often called \hl{appearance probability} (because the flavor $\nu_\beta$ appears).
These probabilities are visualized in \figu{oscprob}:
The mixing angle $\sin^2 2 \theta$ can be interpreted as the \hl{oscillation amplitude}, whereas the mass squared splitting
$\Delta m^2$ can be interpreted as \hl{oscillation frequency}, and its inverse is proportional to the \hl{oscillation length}
$\lambda \equiv (4 \pi E)/\Delta m^2$.
The two probabilities add up to one, which is a consequence of the unitarity of $U$.
Note that even if one introduces an additional CP phase in \equ{utwo}, the quartic invariant $J_{kj}^{\alpha \beta}$ in \equ{osc} cannot become complex for two flavors, which means that there will be no CP violation observable in two-flavor oscillations. For the same reason, \ie, the presence of CP violation in flavor mixing, Kobayashi and Maskawa postulated three flavors in the quark sector, for which they received the Nobel prize 2008.

\subsection{Three-flavor case}

\begin{figure}
\begin{center}
\includegraphics[width=\columnwidth]{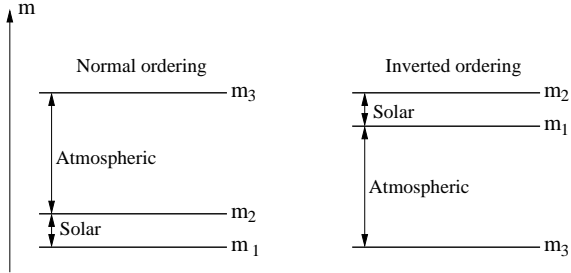}
\end{center}

\vspace*{-1cm}

\caption{\label{fig:spectrum} Neutrino mass eigenstates for normal and inverted mass ordering (not to scale). }
\end{figure}

The current standard assumptions for neutrino oscillations include three active and no sterile neutrino flavors, \ie, $N=3$ and $S=0$. In this case, the possible mass spectra are illustrated in \figu{spectrum} for a \hl{normal} ($\ldm>0$) and \hl{inverted} ($\ldm<0$) \hl{mass ordering}. The splittings between the mass eigenstates are determined by $\sdm \equiv m_2^2-m_1^2$, also called \hl{solar mass splitting}, and $\ldm \equiv m_3^2 - m_1^2$, also called \hl{atmospheric mass splitting} ($\Delta m_{32}^2$ is given by $\ldm - \sdm$). Since the upper bound for the neutrino masses is known to be of the order eV, as we will discuss later, the mass spectrum can be close to this bound, called \hl{degenerate spectrum}, or close to zero, called \hl{hierarchical spectrum}. Since neutrino oscillations are not sensitive to this feature, the terms ``normal/inverted ordering'' and ``normal/inverted hierarchy'' are often used equivalently. The mass ordering and the type of the spectrum are characteristic for neutrino mass models, see \Ref~\cite{Albright:2006cw}. For example, the structure of the neutrino mass matrix is qualitatively different for the normal and inverted ordering in case of a hierarchical spectrum.

The \hl{three-flavor mixing matrix} is, apart from possible Majorana phases not relevant for neutrino oscillations,  typically parameterized as~\cite{PDG}
\begin{eqnarray}
U_{\mathrm{PMNS}} & = & \underbrace{ \left( 
\begin{array}{ccc} 
1 & 0 & 0 \\
0 & c_{23} & s_{23} \\
0 & -s_{23} & c_{23} 
\end{array} \right) }_{\mathrm{Atmospheric \, mixing}} \nonumber \\
& \times &\underbrace{ \left( 
\begin{array}{ccc} 
c_{13} & 0 & s_{13} \, e^{-i \deltacp} \\
0 & 1 & 0 \\
- s_{13} \, e^{i \deltacp} & 0 & c_{13} 
\end{array} \right) }_{\mathrm{Reactor \, mixing}} \nonumber \\
& \times & \underbrace{ \left( 
\begin{array}{ccc} 
c_{12} & s_{12} & 0 \\
-s_{12} & c_{12} & 0 \\
0 & 0 & 1 
\end{array} \right) }_{\mathrm{Solar \, mixing}} \, ,
\label{equ:upmns}
\end{eqnarray}
where $c_{ij} = \cos \theta_{ij}$ and $s_{ij} = \sin \theta_{ij}$. This implies that the neutrino mixing can be parameterized by three mixing angles $\theta_{23}$, $\theta_{13}$, and $\theta_{12}$, which are, for historic reasons, often called atmospheric mixing angle, reactor mixing angle, and solar mixing angle, respectively. In addition, there is one phase $\deltacp$, which leads for $\deltacp \notin \{0,\pi\}$ to CP violation, \cf, \equ{osc}. Note that in this parameterization, $\exp(i \deltacp)$ is multiplied by $\sin \theta_{13}$, which means that a non-zero value of $\theta_{13}$ is required for any measurement of $\deltacp$.

\begin{table*}[t!]\centering
\begin{tabular}{lrrrr}
        \hline
        Parameter & Best-fit & Degrees & 2$\sigma$ & 3$\sigma$ 
        \\
        \hline
        $\Delta m^2_{21}\: [10^{-5}\mathrm{eV^2}]$
        & $7.65^{+0.23}_{-0.20}$  &  & $7.25$--$8.11$ & $7.05$--$8.34$ \\[2mm]
        $|\Delta m^2_{31}|\: [10^{-3}\mathrm{eV^2}]$
        & $2.40^{+0.12}_{-0.11}$  & & $2.18$--$2.64$ & $2.07$--$2.7$5 \\[2mm]
        $\sin^2\theta_{12}$
        & $0.304^{+0.022}_{-0.016}$ & $33^\circ$ & $0.27$--$0.3$5 & $0.25$--$0.37$\\[2mm]
        $\sin^2\theta_{23}$
        & $0.50^{+0.07}_{-0.06}$ & $45^\circ$ & $0.39$--$0.63$ & $0.36$--$0.67$\\[2mm]
        $\sin^2\theta_{13}$
        & $0.01^{+0.016}_{-0.011}$  & $6^\circ$ & $\leq$ $0.040$ & $\leq$ $0.056$ \\[2mm]
        $\deltacp$ & \multicolumn{4}{l}{Currently no information} \\
        \hline
\end{tabular}
\caption{ \label{tab:summary} Current best-fit values with 1$\sigma$ errors,
  best-fit values in degrees (angles only), and 2$\sigma$ and 3$\sigma$ intervals 
(1 d.o.f.) for the three-flavor neutrino oscillation parameters from global data; 
  adopted from \Ref~\cite{Schwetz:2008er}.}
\end{table*}

Together with the two independent mass squared differences $\sdm$ and $\ldm$, we have six neutrino oscillation parameters. The mixing matrix can be fully described by mixing angles in the parameter ranges $ \theta_{ij} \in [0, \pi/2]$ and $\deltacp \in [0, 2 \pi[$ (see, \eg, \Ref~\cite{deGouvea:2008nm}). If the neutrinos are Majorana particles, the mixing matrix should be replaced by
\begin{equation}
U_{\mathrm{PMNS}} \rightarrow U_{\mathrm{PMNS}} \times \mathrm{diag}(1, e^{i \alpha}, e^{i \beta}) \, , \label{equ:maj} 
\end{equation}
because these additional phases cannot be absorbed in a re-definition of the Majorana fields. These phases with the physical parameter ranges $[0, \pi]$ enter the description of $0\nu\beta\beta$ decay, but not into neutrino oscillations. They are called \hl{Majorana phases}, whereas $\deltacp$ is often referred to as \hl{Dirac CP phase} (meaning that it is also present for Dirac neutrinos).
Note that the parameterization in \equ{upmns} is somehow arbitrary. It only makes sense in combination or comparison with the quark sector, where $V_{\mathrm{CKM}}$  describes the rotation between the up- and down-type quark states using the {\em same} parameterization. For example, one may test a possible connection between the quark and lepton sectors, such as by a unifying theory, or obtain hints for the generation of the flavor structure, which may or may not have the same origin in both sectors.

The current knowledge on the three-flavor neutrino oscillation parameters is summarized in \Tab~\ref{tab:summary}.
We can read off two qualitative observations from this table, which will be relevant for our analytical treatment:
\begin{enumerate}
\item
 One of these mass squared differences is much smaller than the other two: $\sdm \ll \ldm \simeq \Delta m_{32}^2$. This leads to a hierarchy of the neutrino mass splittings, as illustrated in \figu{spectrum}.
\item
 Two of the mixing angles, $\theta_{23}$ and $\theta_{12}$, are very large, whereas one mixing angle $\theta_{13}$ is small -- at most of the size of the Cabibbo angle $\theta_C$ in the quark sector.\footnote{Recently, a $1.6\sigma$ claim for $\theta_{13}>0$ has been made from the global analysis of all oscillation data~\cite{Fogli:2008jx}. However, this claim depends on details of the analysis and may very well come from statistical fluctuations, see \Refs~\cite{Maltoni:2008ka,GonzalezGarcia:2010er} for a more detailed discussion.} 
\end{enumerate}
There might be even  \hl{maximal mixing} $\theta_{23} = \pi/4$, which could point (possibly with a vanishing $\theta_{13}$) towards a fundamental symmetry between $\nu_\mu$ and $\nu_\tau$.\footnote{Maximal mixing is, from the oscillation point of view, illustrated in \figu{oscprob}. In this case, the two-flavor survival probability may even vanish at certain $L$ and $E$.} For $\theta_{12}$, maximal mixing is excluded at more than $5 \sigma$. However, the mixing angles are compatible with the so-called tri-bimaximal mixing~\cite{Harrison:2002er}, where $\sin^2 \theta_{12} = 1/3$, $\sin^2 \theta_{23}=1/2$, and $\sin^2 \theta_{13}=0$
leading to a specific form of the neutrino mass matrix, which has been motivated by a large class of models. 
From \Tab~\ref{tab:summary}, one can also read off the primary remaining quantities of interest for future experiments:
\begin{itemize}
\item
 The value of $\theta_{13}$, and if it is non-zero.
\item
 The sign of $\ldm$, \ie, the ordering of the neutrino masses.
\item 
 The value of $\deltacp$ (only if $\theta_{13}>0$), and if it is violating the CP symmetry, \ie, if $\deltacp \notin \{0, \pi\}$.
\item 
 The exact value of $\theta_{23}$, in particular, if maximal mixing $\theta_{23}=\pi/4$ can be excluded, and if $\theta_{23}>\pi/4$ or $<\pi/4$, the $\theta_{23}$ \hl{octant}. 
\end{itemize}
Above we have mentioned that there is currently no evidence for additional sterile neutrino species or other new physics contributing to neutrino oscillations in a leading role. For example, the evidence for active-sterile neutrino oscillations from the LSND experiment~\cite{Aguilar:2001ty} has been ruled out~\cite{Maltoni:2007zf}. This, however, does not mean that it is not interesting to look for sub-leading new physics effects in neutrino oscillations, since particular classes of effects might be primarily visible there.

\subsection{Two-flavor sub-sectors}

In the following, let us use the qualitative knowledge on the neutrino oscillation parameters in order to reconstruct the different two-flavor sub-sectors which have lead to the current knowledge. This is not meant to be a complete review, but only a short discussion to give the reader some idea of the relevant measurements. For the sake of simplicity, let us first of all assume that $\sdm \ll \ldm$ and $U_{\mathrm{PMNS}}$ is real. Then we have from \equ{osc}
\begin{eqnarray}
 P_{\alpha \beta} & = & \delta_{\alpha \beta} - 4 \, \left( J_{31}^{\alpha \beta} + J_{32}^{\alpha \beta} \right) \sin^2 \Delta_{31} \nonumber \\
 & - & 4 \, J_{21}^{\alpha \beta} \sin^2 \Delta_{21} \, ,\label{equ:psimple}
\end{eqnarray}
where $\Delta_{ij} \equiv \Delta m_{ij}^2 \, L/(4 E)$. We can now choose one of the following two oscillation frequencies: 
\begin{description}
\item[The atmospheric oscillation frequency] or $\Delta_{31} \simeq \pi/2$. This necessarily leads to $\Delta_{21} \ll 1$, \ie, the second oscillatory  part in \equ{psimple} is very small.
\item[The solar oscillation frequency] or $\Delta_{21} \simeq \pi/2$. This necessarily leads to $\Delta_{31} \gg 1$,  meaning that the first oscillatory part in \equ{psimple} averages out; see also \equ{average}.
\end{description}
Note that we ``choose'' an oscillation frequency by the neutrino energy $E$, determined by the neutrino source, and the baseline $L$, determined by the experimental configuration. In addition, note that the names ``solar'' and ``atmospheric'' frequency or mass squared splitting (if referring to the corresponding $\Delta m^2$) has again historical reasons, as we shall see below.
In the limit $\theta_{13} \rightarrow 0$,  the mixing matrix in \equ{upmns} simplifies to 
\begin{equation}
U_{\mathrm{PMNS}}^{\theta_{13} \rightarrow 0} = \left( 
\begin{array}{ccc}
c_{12} & s_{12} & 0 \\
-s_{12} \, c_{23} & c_{12} \, c_{23} & s_{23} \\
s_{12} \, s_{23} & - c_{12} \, s_{23} & c_{23} 
\end{array}
\right) \, .
\label{equ:usimple}
\end{equation}
Using Eqs.~(\ref{equ:psimple}) and~(\ref{equ:usimple}), one easily obtains the leading order two-flavor oscillation probabilities of the following experiment classes, which have been carried out so far:

{\bf Atmospheric experiments} use the neutrinos produced (mainly) by pions as secondaries of cosmic ray interactions in the Earth's atmosphere. Charged pion decays produce fluxes of electron and muon neutrinos (and anti-neutrinos), see \equ{pion}. The detectors, such as Super-Kamiokande~\cite{Fukuda:1998mi}, can detect electron or muon neutrinos, which means that the following four oscillation probabilities are interesting:
\begin{eqnarray}
P_{ee} & \simeq & 1 \, ,  \quad P_{e \mu} \simeq P_{\mu e} \simeq 0 \nonumber \\
P_{\mu \mu} & \simeq & 1 - \sin^2 2 \theta_{23} \, \sin^2 \Delta_{31} \, . \label{equ:patm}
\end{eqnarray}
Obviously, atmospheric neutrino oscillations can, to leading order, be described by the two-flavor limit with the parameters $\theta_{23}$ and $\Delta m_{31}^2$ (in fact, the neutrinos change flavor into $\nu_\tau$, which are invisible to the detector). Therefore, these oscillation parameters are often called \hl{atmospheric parameters}.

{\bf Solar experiments} historically detect the neutrinos produced by fusion reactions in the Sun, which cannot be described by the vacuum oscillation framework we have introduced so far. However, we can describe a very long baseline reactor experiment using multiple nuclear power plants in Japan as neutrino sources: the KamLAND experiment (``Kamioka Liquid scintillator Anti-Neutrino Detector'')~\cite{Araki:2004mb}.
Since nuclear reactors produce $\bar\nu_e$ only (by beta decays), which might be detected by inverse beta decays, the applicable oscillation probability from Eqs.~(\ref{equ:psimple}) and~(\ref{equ:usimple}) is
\begin{equation}
P_{\bar e \bar e} \simeq  1 - \sin^2 2 \theta_{12}  \, \sin^2 \Delta_{21} \, .
\label{equ:rlong}
\end{equation}
The parameters measured in this experiment are the \hl{solar parameters}, and the probability again corresponds to the two-flavor limit. In fact, here the $\bar\nu_e$ oscillate into a superposition of about equal amounts of $\bar\nu_\mu$ and $\bar\nu_\tau$ (with their ratio determined by $\theta_{23}$).

\begin{figure}
\begin{center}
\includegraphics[width=\columnwidth]{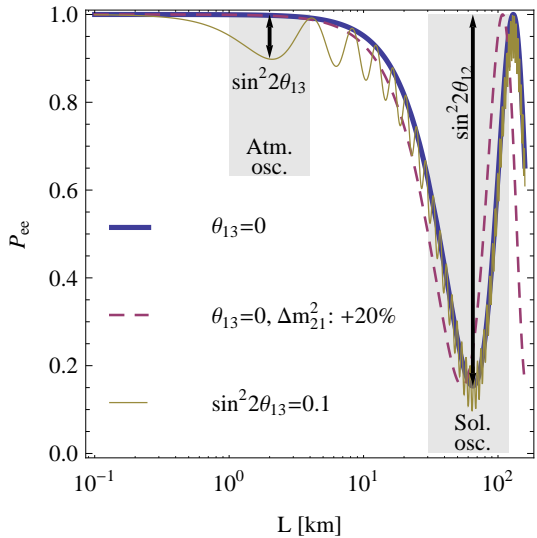}
\end{center}
\vspace*{-1cm}
\caption{\label{fig:reactor} Reactor electron antineutrino disappearance probability, illustrated for different sets of parameters and perfect resolution. }
\end{figure}

{\bf Reactor experiments for $\boldsymbol{\theta_{13}}$.} 
Relaxing the condition $\theta_{13} \rightarrow 0$, \ie, using \equ{upmns} instead of \equ{usimple}, one obtains different from \equ{rlong}
\begin{eqnarray}
P_{\bar e \bar e} & \simeq  & 1 -  \sin^2 2 \theta_{13}  \, \sin^2 \Delta_{31} \nonumber \\
& - & \cos^4 \theta_{13} \, \sin^2 2 \theta_{12}  \, \sin^2 \Delta_{21} \, .
\label{equ:rall}
\end{eqnarray}
Choosing a much shorter baseline than for the solar reactor experiments above, one selects the atmospheric oscillation frequency in order to obtain $P_{\bar e \bar e} \simeq  1 -  \sin^2 2 \theta_{13}  \, \sin^2 \Delta_{31}$, with small deviations to be interpreted as a signal for a non-zero $\theta_{13}$. Therefore, $\theta_{13}$ is also often called \hl{reactor angle}. An example for a corresponding experiment has been the CHOOZ experiment (named after its site)~\cite{Apollonio:1999ae}. 
The interplay between atmospheric and solar oscillation frequency is illustrated in \figu{reactor}. The thick solid curve corresponds to $\theta_{13}=0$, \ie, \equ{rlong}. In this case, the oscillation dip is at about 60~km in the gray-shaded solar oscillation window. Small changes in $\sdm$ will move this dip, as illustrated by the dashed curve by increasing $\sdm$ by 20\%, which means that such an experiment is very sensitive to $\sdm$. If $\theta_{13}>0$ (thin solid curve) the faster atmospheric oscillation will be superimposed, \cf, \equ{rall}, leading to sensitivity to $\theta_{13}$ in the atmospheric oscillation window at about 1-2~km. At the longer baselines, the atmospheric oscillations can in practice not be resolved and are averaged out.

{\bf Conventional neutrino beams.} In this case, the neutrinos are produced (mainly) by pion decays such as in the atmosphere, but using a man-made neutrino source. They are detected as electron and muon flavors, such as in the currently running MINOS experiment (``Main Injector Neutrino Oscillation Search'', Fermilab)~\cite{Michael:2006rx}, or as tau neutrinos, such as in the OPERA experiment (``Oscillation Project with Emulsion-tRacking Apparatus'') in the CNGS (``CERN to Gran Sasso'') beam~\cite{Duchesneau:2002yq}. The probabilities of interest are the same as in \equ{patm} (except for OPERA, where $P_{\mu \tau} \simeq 1- P_{\mu \mu}$), for example, MINOS has provided an improved measurement of $\ldm$. However, with the increasing statistics of such experiments, corrections from $P_{\mu e} \simeq 0$, as in \equ{patm}, can be measured. As we will demonstrate later, these corrections are not only a measure of $\theta_{13}$, but also contain the information necessary to extract CP violation. 

In summary, neutrino oscillations have so far mostly been derived from two-flavor sub-sectors of the general three-flavor framework. Depending on the experiment, the two-flavor probabilities in \equ{twoflavor} are described by different sets of parameters, such as $\{ \Delta m_{31}^2, \theta_{23} \}$ (atmospheric parameters) for the atmospheric experiments,  $\{ \Delta m_{21}^2, \theta_{12} \}$ (solar parameters) for the long baseline reactor experiments, and $\{ \Delta m_{31}^2, \theta_{13} \}$ for the short baseline reactor experiments;
see also \figu{oscprob} for typical values of the mixing angles (\cf, \Tab~\ref{tab:summary}).
 Especially the measurement of $\deltacp$ will be a direct test of the three-flavorness of neutrino oscillations, as we shall see in the next section.

\section{Future precision oscillation physics}
\label{sec:prec}

In this section, we discuss neutrino oscillations beyond the two-flavor sub-sector measurements, which have lead to the current knowledge. We introduce matter effects in Earth matter to neutrino oscillations, and we show how three-flavor effects can be accessed. Furthermore, we introduce future experiment classes and discuss their simulation. Finally, in the era of precision neutrino physics, we also show examples of interesting new physics effects, and how they can be tested.

\subsection{Matter effects in neutrino oscillations}

In order to discuss future precision neutrino oscillation physics, we need another key ingredient of neutrino oscillations, which is the matter effect~\cite{Mikheev:1985gs,Mikheev:1986wj,Wolfenstein:1978ue}. This effect implies that coherent forward scattering in matter by charged current and neutral current interactions affects neutrino oscillations. Neutral current interactions occur for all (active) flavors, leading to an overall phase which can be subtracted, whereas charged current interactions are only possible for electron neutrinos (or anti-neutrinos). The reason for this asymmetry is that ordinary matter consists of electrons, protons, and neutrons, whereas there are no muons and tauons required as SU(2) counterparts of the $\nu_\mu$ and $\nu_\tau$ for charged current interactions. This leads to an effective net potential $A_{\mathrm{CC}}$ on the electron flavor, which can in flavor space be described as a term adding to \equ{ham} in the electron flavor\footnote{Note that, compared to \equ{ham}, we have used already the ultra-relativistic approximation \equ{ur} here, and we have subtracted an overall phase factor.}:
\begin{eqnarray}
\mathcal{H}_F & = & U \, \left( 
\begin{array}{ccc}
0 & 0 & 0 \\
0 & \frac{\sdm}{2E} & 0 \\
0 & 0 & \frac{\ldm}{2E} 
\end{array}
\right) \, U^\dagger  \nonumber  \\
& + & \left( 
\begin{array}{ccc}
V_{\mathrm{CC}} & 0 & 0  \\
0 & 0 & 0 \\
0 & 0 & 0 
\end{array}
\right) \, . 
\label{equ:ham3}
\end{eqnarray}
Here $V_{\mathrm{CC}}= \pm \sqrt{2} \, G_F \, n_e$ is the \hl{matter potential} with $n_e \simeq Y_e \, \rho/m_N \simeq \rho/(2 m_N)$ the electron density in matter. The quantity $Y_e$ describes the number of electrons per nucleon with the nucleon mass $m_N$. Furthermore, the sign in $V_{\mathrm{CC}}$ is positive (negative) for neutrinos (anti-neutrinos).
The evaluation of Eq.~(\ref{equ:ham}a) with \equ{ham3} is straightforward if the Hamiltonian is not explicitely time-dependent. In this case, one simply re-diagonalizes \equ{ham3} in order to obtain the mixing matrix and mass eigenstates in matter (see, \eg, \Ref~\cite{Winter:2006vg}). This approach, is, in fact, often used in numerical calculations, where one typically
 evaluates the Hamiltonian in layers of constant matter density. Analytical computations, on the other hand, are relatively simple in the two-flavor limit in both constant and (slowly enough) varying matter densities.  In the Sun and in supernovae, where matter effects are especially important because of the extremely high electron densities, one has to deal with varying matter densities. Here we first concentrate on the simpler case of Earth matter. As long as the baseline does not cross the Earth's core, \ie, $L \lesssim 10 \, 700 \, \mathrm{km}$, using a constant matter density is a good first order approximation. In the two-flavor limit, we obtain from \equ{ham3} by multiplying out and subtracting a global phase\footnote{Adding or subtracting to $\mathcal{H}_F$ a quantity proportional to the unit matrix leaves the oscillation physics unchanged.}
{\small
\begin{equation}
\mathcal{H}=\frac{1}{4E} \left( 
\begin{array}{cc}
-\Delta m^2 \cos 2 \theta + A_{\mathrm{CC}} & \Delta m^2 \sin 2 \theta \\
 \Delta m^2 \sin 2 \theta & \Delta m^2 \cos 2 \theta - A_{\mathrm{CC}}
\end{array}
\right)
\label{equ:hmatter}
\end{equation}
} % small
with
\begin{equation}
A_{\mathrm{CC}} = 2 \, E \, V_{\mathrm{CC}} = \pm 2 \sqrt{2} \, E\, G_F \, n_e \, .
\end{equation}
Compared to vacuum, where
\begin{equation}
\mathcal{H}=\frac{1}{4E} \left( 
\begin{array}{cc}
-\Delta m^2 \cos 2 \theta  & \Delta m^2 \sin 2 \theta \\
 \Delta m^2 \sin 2 \theta & \Delta m^2 \cos 2 \theta 
\end{array}
\right) \, ,
\label{equ:hvac}
\end{equation}
we can use the same form in matter using effective parameters $\Delta \tilde m^2$ and $\tilde \theta$
\begin{equation}
\mathcal{H}=\frac{1}{4E} \left( 
\begin{array}{cc}
-\Delta \tilde m^2 \cos 2 \tilde \theta  & \Delta \tilde m^2 \sin 2 \tilde \theta \\
 \Delta \tilde m^2 \sin 2 \tilde \theta & \Delta \tilde m^2 \cos 2 \tilde \theta 
\end{array}
\right) \, ,
\label{equ:hmatter2}
\end{equation}
leading to the same form of the oscillation probabilities \equ{twoflavor}:
\begin{eqnarray}
P_{\alpha \alpha} & = & 1 - \sin^2  2 \tilde\theta  \,  \sin^2 \left( \frac{ \Delta \tilde m^2 L}{4 E} \right) \, , \nonumber \\
P_{\alpha \beta} & = & \sin^2  2 \tilde\theta  \,  \sin^2 \left( \frac{ \Delta \tilde m^2 L}{4 E} \right) \, ,
\label{equ:twoflavorm}
\end{eqnarray}
From the comparison between \equ{hmatter2} and \equ{hmatter}, one can easily show that the parameter mapping is
\begin{eqnarray}
\Delta \tilde m^2 & = & \Delta m^2 \, \xi \, , \quad \sin 2 \tilde \theta = \frac{\sin 2 \theta}{\xi} \, , \nonumber \\ 
\xi & = & \sqrt{\left( \cos 2 \theta - \hat{A} \right)^2 + \sin^2 2 \theta } \, ,
\label{equ:mapping}
\end{eqnarray}
with
\begin{equation}
 \hat{A} = \frac{A_{\mathrm{CC}}}{\Delta m^2} = \pm \frac{2 \, \sqrt{2} \, E \, G_F \, n_e }{\Delta m^2} \, . \label{equ:ahat}
\end{equation}

These formulas are useful to demonstrate a number of important consequences. First of all, in the limit $\Delta \tilde m^2 L/(4 E) \ll 1$, the oscillating term in \equ{twoflavorm} can be expanded, and one immediately can read off these equations that the $\xi$ parameters cancel, and the vacuum probabilities are recovered. This means that long enough baselines are relevant to observe matter effects. Second, one can read off from \equ{mapping} that the oscillation angle (or amplitude) becomes resonantly enhanced for $\cos 2 \theta = \hat{A}$. The corresponding resonance energy is given by
\begin{equation}
E_{\mathrm{res}} \, [\mathrm{GeV}] \sim 13 \, 200 \, \cos 2 \theta \, \frac{\Delta m^2 \, [\mathrm{eV}^2]}{\rho \, [\mathrm{g/cm^3}]} \, .
\label{equ:eres}
\end{equation}
For $\rho = 3.4 \, \mathrm{g/cm^3}$ (the average density for $L=4 \, 000 \, \mathrm{km}$), $\Delta m^2 = \ldm$ from \Tab~\ref{tab:summary}, and $\theta=\theta_{13} \simeq 0$, this evaluates to a \hl{resonance energy} of $E_{\mathrm{res}} \simeq 9.3 \, \mathrm{GeV}$.
Therefore, relatively high neutrino energies are required for substantial Earth matter effects. Third, the resonance will only occur for $\mathrm{sgn}(\hat{A}) = +1$, whereas for $\mathrm{sgn}(\hat{A}) = -1$ there will be an \hl{antiresonance} with suppressed oscillation amplitude; \cf, \equ{ahat}. This implies that a resonant transition occurs for neutrinos and $\Delta m^2>0$, or anti-neutrinos and $\Delta m^2<0$. In fact, one can use this effect to measure the mass ordering  with high sensitivity, because for strong matter effects the rates will be strongly affected by the sign of $\Delta m^2$.
And fourth, we have learned above that the two-flavor probabilities should be CP invariant, whereas they are not in the presence of matter, \ie, in \equ{twoflavorm}. For a CP invariant problem in matter, one would have to CP-conjugate the matter potential as well, which means that the Earth matter would have to be replaced by antimatter (which is, of course, impossible). Therefore, matter effects violate the CP (and even CPT) symmetries in an extrinsic form. In any realistic experiment with strong matter effects, the CP violation has therefore to be extracted from a convolution of the \hl{intrinsic} (from $\deltacp$) and \hl{extrinsic} (from the matter potential) \hl{CP violation}, which implies that \equ{cpasym} is not a very good description of CP violation in neutrino oscillations if matter effects are present.

\subsection{Three-flavor effects}

For the illustration of three-flavor effects, the most relevant oscillation channels will be the $\nu_\mu \rightarrow \nu_e$ (or $\nu_e \rightarrow \nu_\mu$) channels. In \equ{patm}, we have learned that $P_{e \mu} \simeq P_{\mu e} \simeq 0$ for atmospheric experiments to a first approximation, which means that deviations from zero will be driven by $\theta_{13}$ and the solar oscillation contribution. In order to switch these effects on, it is therefore appropriate to expand these appearance probabilities to second order in $\sin 2 \theta_{13}$ and the hierarchy parameter $\alpha \equiv \sdm/\ldm \simeq 0.03$ as~\cite{Cervera:2000kp,Freund:2001pn,Akhmedov:2004ny}
\begin{eqnarray}
 P_{e\mu} &\simeq& 
 \sin^22\theta_{13} \sin^2\theta_{23} 
\frac{\sin^2[(1-\hat{A})\Delta_{31}]}{(1-\hat{A})^2}\nonumber \\
&\pm& \alpha \sin2\theta_{13} \sin\deltacp \sin2\theta_{12} \sin2\theta_{23}  \nonumber \\
 & & \times  \sin(\Delta_{31}) \frac{\sin(\hat{A}\Delta_{31})}{\hat{A}}
\frac{\sin[(1-\hat{A})\Delta_{31}]}{(1-\hat{A})} \nonumber \\
&+& \alpha \sin2\theta_{13} \cos\deltacp \sin2\theta_{12} \sin2\theta_{23} \nonumber \\
 & & \times \cos(\Delta_{31}) \frac{\sin(\hat{A}\Delta_{31})}{\hat{A}}
\frac{\sin[(1-\hat{A})\Delta_{31}]}{(1-\hat{A})} \nonumber \\
&+& \alpha^2 \cos^2\theta_{23} \sin^22\theta_{12} 
\frac{\sin^2(\hat{A}\Delta_{31})}{{\hat{A}}^2} \, .
\label{equ:papp}
\end{eqnarray}
Here the sign of the second term refers to neutrinos (plus) or anti-neutrinos (minus). 
Note that the sign of $\hat{A}$, defined in \equ{ahat}, depends on 
neutrinos or anti-neutrinos as well. The T-inverted probability
$P_{\mu e}$, however, can be obtained from \equ{papp} by changing the
sign of the second term only.

From \equ{papp}, we can immediately read off that all of the interesting
quantities $\theta_{13}$, the mass hierarchy (by the effect in $\hat{A}$),
and $\deltacp$ (by the second and third terms) can be measured in principle
if the spectral dependence of the different terms can be disentangled.
However, because of the complex parameter dependence and matter effects, continuous \hl{correlations} and several discrete
\hl{degeneracies} remain in the parameter space even if both neutrinos
and anti-neutrinos are used: The
$(\delta, \theta_{13})$~\cite{Burguet-Castell:2001ez},
$\mathrm{sgn}(\Delta m_{31}^2)$~\cite{Minakata:2001qm}, and
$(\theta_{23},\pi/2-\theta_{23})$~\cite{Fogli:1996pv} degeneracies,
\ie, and overall ``eight-fold'' degeneracy~\cite{Barger:2001yr}. 
Using enough energy resolution, different baselines, different oscillation
channels, or more statistics, the correlations and degeneracies can be resolved.
One example is the condition $\sin( \hat{A} \Delta_{31} ) = 0$ in \equ{papp},
which makes the second to fourth terms disappear, and leads to a  clean measurement of $\theta_{13}$ and the mass hierarchy.
This condition evaluates to $\sqrt{2} \, G_F \, n_e \, L = 2 \pi$ independent of
neutrino energy and oscillation parameters, or $L \simeq 7 \, 500 \, \mathrm{km}$,
the so-called \hl{magic baseline}~\cite{Huber:2003ak}.
It is in general a good strategy to
combine a shorter baseline with weaker matter effects in order to measure CP violation 
with a longer baseline with stronger matter effects to measure the mass hierarchy.

\begin{figure}[t!]
\begin{center}
\includegraphics[width=0.8\columnwidth]{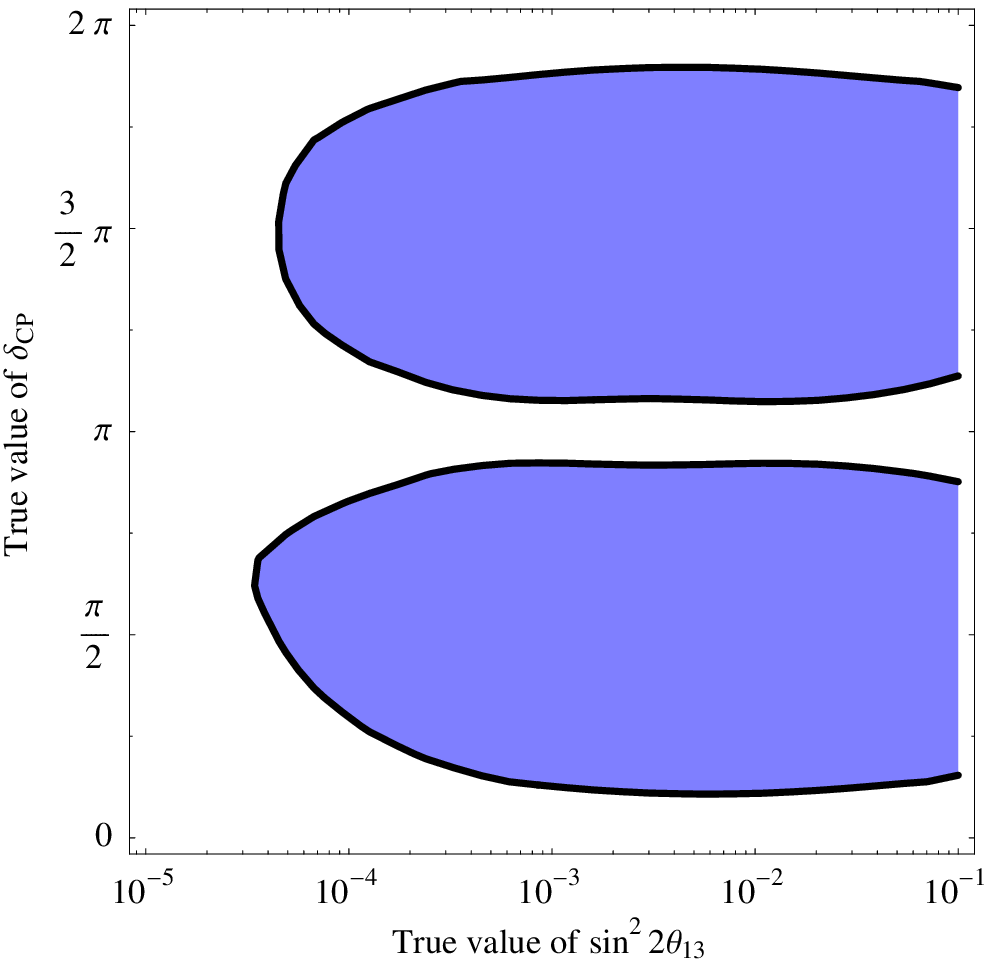} 

\hspace*{0.2cm}\includegraphics[width=0.8\columnwidth]{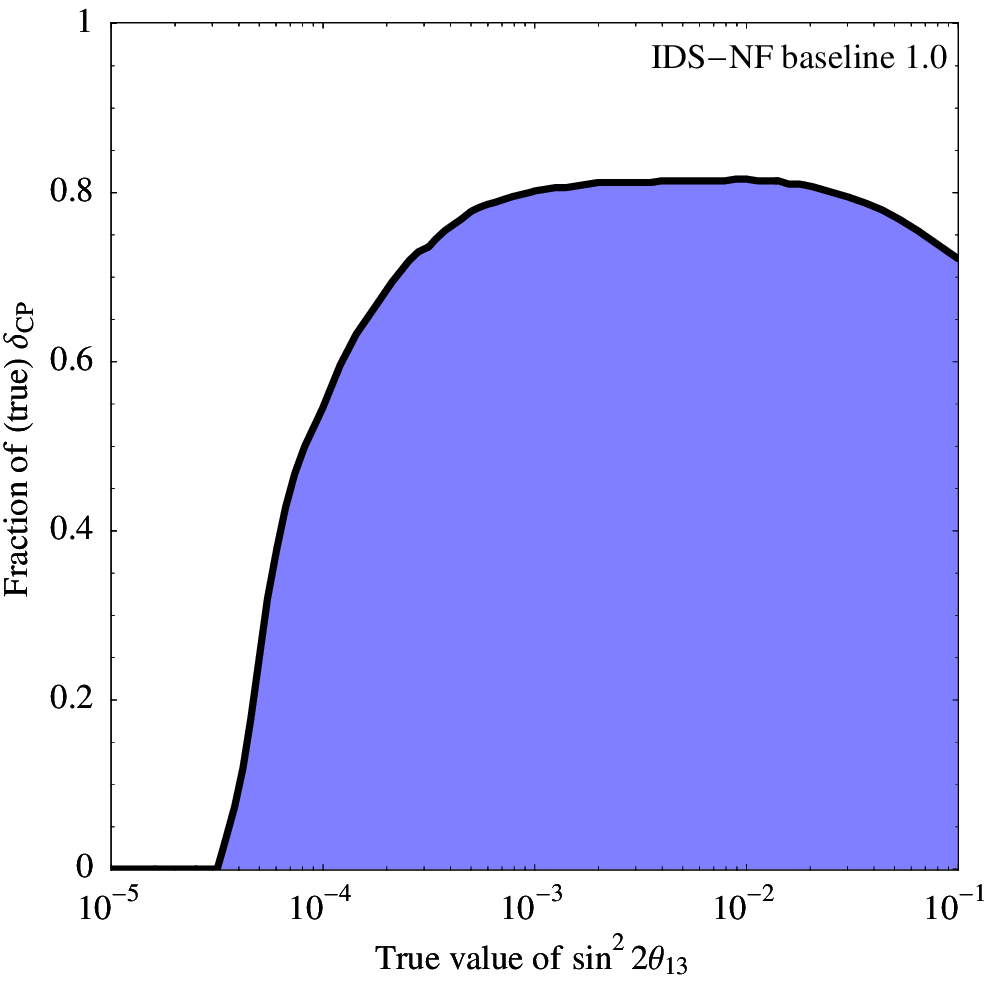}
\end{center}
\vspace*{-1cm}
\caption{\label{fig:cpdid} Discovery reach for CP violation ($3 \sigma$) for the IDS-NF neutrino factory~\cite{ids} as a function of (true) $\stheta$ and (true) $\deltacp$ (upper panel) or fraction of (true) $\deltacp$ (lower panel). Normal mass ordering assumed.}
\end{figure}

Let us take a closer look at the discovery reach for CP violation. A discovery of CP violation will be made if all CP conserving solutions $\deltacp=0$ and $\pi$ can be excluded at a certain confidence level for an arbitrary (allowed) choice of the other oscillation parameters. In practice, one marginalized over the other parameters. From \Tab~\ref{tab:summary}, we know that the solar and atmospheric parameters are already very well known, whereas $\stheta \lesssim 0.1$ and $\deltacp$ are unknown quantities. The performance of any future experiment will however depend on $\stheta$ and $\deltacp$ within their allowed ranges, \ie, the values which Nature has actually implemented. These values are often referred to as \hl{true values} or \hl{simulated values},  and correspond to data in an existing experiment. For the CP violation measurement, $\stheta$ and $\deltacp$ (and the mass ordering) are the critical parameters which determine the actual experiment performance. Consequently, any future experiment should operate in a (true) $\theta_{13}$ and $\deltacp$ range as large as possible. For the quantification of the CP violation performance one therefore often shows the region in $\stheta$ and $\deltacp$ where CP violation will be discovered, as illustrated in \figu{cpdid} (upper panel) for a neutrino factory. Obviously, if $\stheta$ is too small, the second and third terms in \equ{papp} will be too small, and no CP violation will be discovered. If $\deltacp$ is too close to one of the CP conserving values, CP violation cannot be discovered either. A different representation of the CP violation discovery potential is shown in the lower panel of \figu{cpdid}. In this case, for each $\stheta$, the sensitive regions are stacked, and the fraction of $\deltacp$ for which CP violation will be discovered is shown. For example, for $\stheta=10^{-3}$, CP violation will be discovered for about 80\% of all possible values of $\deltacp$. This representation turns out to be useful for experiment performance comparisons, see, \eg, \Ref~\cite{Bandyopadhyay:2007kx}. Note that in \figu{cpdid}, the combination of 4000~km and 7500~km baselines is used to resolve the degeneracies and to disentangle the intrinsic from the extrinsic (matter effect)  CP violation, which would otherwise lead to irregularities in the discovery regions.

The mass ordering is mainly measured by the first term in \equ{papp}, which means that the sensitivity depends on $\stheta$.  For the sake of simplicity, let us chose the magic baseline, where only this first term survives. Then \equ{papp} reduces to the two-flavor limit in \equ{twoflavorm}, apart from the factor $\sin^2 \theta_{23}$. In this case, for small enough $\theta_{13}$, the parameter mapping in \equ{mapping} is given by $\xi \simeq 1 - \hat{A}$, and the resonance condition corresponds to $\hat{A} \rightarrow 1$. At the resonance, $P_{e \mu} \propto \Delta^2 \propto L^2$, which compensates for the $1/L^2$ flux drop of the event rates. Therefore, the event rates stay almost constant for a wide baseline range; \cf, Fig.~1 in \Ref~\cite{Freund:1999gy}. Since the effect is opposite for the anti-resonance (such as the other mass ordering), long baselines and high enough neutrino energies covering the matter resonance energy are beneficial  for the discrimination of the mass ordering.

\subsection{Future experiment classes}

Here we consider future reactor and accelerator-based \hl{long baseline experiments} to find the unknown quantities,  $\theta_{13}$, the mass ordering, and CP violation.  The experiments are typically classified by their neutrino production mechanism:

{\bf Reactor experiments with two detectors} use neutrinos produced by beta decay in nuclear fission reactors, such as nuclear power plants. They are similar to the reactor experiments from the last section, measuring $\theta_{13}$ in the short baseline limit of \equ{rall}. As a major improvement of the CHOOZ experiment, additional near detectors help to better control systematics, such as the normalization of the flux. Examples for these experiments will be Double Chooz~\cite{Ardellier:2006mn} and Daya Bay~\cite{Guo:2007ug}.

{\bf Superbeams} follow the technology of conventional beams producing neutrinos by (mostly) pion decays. Compared to the conventional beams, the proton beam intensity on the target will be higher, and the detectors will be more massive. In addition, the \hl{off-axis technology}~\cite{offaxis} is typically used, which means that the main detector is placed slightly off the main beam axis to reduce the beam energy and to over-proportionally reduce backgrounds intrinsic to the beam. Examples are the currently planned T2K (``Tokai to Kamioka'')~\cite{Itow:2001ee} and NO$\nu$A  (``NuMI Off-axis Neutrino Appearance'')~\cite{Ayres:2004js} experiments.

{\bf Superbeam upgrades} are more speculative ideas to push the conventional technology to its limits. This includes extremely high thermal powers in the target, and detector masses in the megaton class (for water Cherenkov detectors). One typically distinguishes two categories: \hl{narrow band beams} are based on the off-axis technology, whereas \hl{wide band beams} are using a detector operated on the beam axis. Note that ``narrow'' and ``wide'' refer to the broadth of the energy spectrum here. There are now many ideas under discussion and evaluation. Examples for narrow band beams are upgrades for the T2K experiment using a megaton-size water Cherenkov detector (T2HK -- ``Tokai to Hyper-Kamiokande'')~\cite{Itow:2001ee}), or even splitting this detector mass between sites in Japan and Korea (T2KK -- ``Tokai to Kamioka and Korea'')~\cite{Ishitsuka:2005qi}). An example for a wide band beam is an on-axis beam from Fermilab (USA) to an Deep Underground Science and Engineering Laboratory (DUSEL) in the Homestake mine (South Dakota, USA)~\cite{Barger:2007yw}.

{\bf Beta beams} produce neutrinos by beta decays of boosted isotopes in straight sections of a storage ring~\cite{Zucchelli:2002sa}. Compared to the superbeams, one has a flavor-clean electron neutrino beam
with very predictable spectrum. However, the ion source has to produce enough radioactive ions per time frame,
and a relatively large accelerator has to boost them to their target energies.
This approach is currently under study from both the experimental and theoretical point of view, such as within the EURISOL (``European Isotope Separation On-Line'') design study~\cite{EURISOL}. 

{\bf Neutrino factories}  produce neutrinos by muon (anti-muon) decays in straight sections of a storage ring~\cite{Cervera:2000kp,Geer:1998iz,DeRujula:1998hd,Barger:1999fs}.
In this case, the spectrum from the purely leptonic muon (anti-muon) decays is very well known, but not flavor clean:
 Since both $\nu_\mu$ ($\bar\nu_\mu$) and $\bar\nu_e$ ($\nu_e$) are produced simultaneously in equal amounts (\cf, seond line of \equ{pion}), charge identification of the secondaries is required in the detector to distinguish the original flavors. In principle, the muon production is technologically straightforward, but the muons have to be collected, cooled\footnote{Here ``cooling'' refers to producing bunches small enough in phase space to be acceptable for the accelerator.}, and accelerated fast enough before they decay. 
This technology is currently under investigation in the international design study for the neutrino factory IDS-NF~\cite{ids}. Very interestingly, a neutrino factory uses in part the same technology required by a muon collider, which means that it could be a first step towards such an experiment. The CP violation discovery reach of the IDS-NF neutrino factory is shown in \figu{cpdid}. CP violation could be measured more than three orders of magnitude in the oscillation amplitude $\stheta$ beyond the current bound.

\subsection{Simulation of future experiments}

Typical questions regarding the optimization of future experiments
concern the {\em status quo} at the time of the decision, the timescales
of different experiments, the comparison of experiments, and the complementarity
of the information obtained. 

As far as the optimization of individual experiment classes is concerned,
one has to distinguish between green-field scenarios and site specific proposals.
The first can, in principle, be attached to any major high energy
laboratory (with possibly substantial extra effort), whereas the latter depend on a specific
site, where some components may be already available. The study of green-field
scenarios is especially useful to identify the optimal setups from the physics
point of view, and to quantify the tradeoff for specific sites.
Of course, the further in the future an experiment is, the more green-field
the considered scenarios will be.

As far as the quantities of interest for the optimization are concerned, there are
typically:
\begin{enumerate}
\item
 The energy of the parent particles, such as the
\begin{itemize}
\item
Ions for a beta beam (quantified by the boost factor $\gamma$) 
\item
Muons for a neutrino factory (quantified by the muon energy $E_\mu$)
\item
Pions/kaons for a conventional beam (typically quantified by the proton energy $E_p$, where the pions and kaons are produced by the interactions of the protons with a solid target).
\end{itemize}
 \item 
 The baseline $L$, \ie, the distance between source and detector.
\item 
 The integrated luminosity, which is proportional to the \\
number of useful parent decays $\times$ running time $\times$ detector mass.
\item
 Detector properties, such as efficiencies, energy resolution, and the ability to measure the
charge of the secondary particle.
\item
 Systematical errors (and their treatment).
\item
 Different parent particles used (such as different isotopes for beta beams).
\item
 The addition of other oscillation channels.
\item
 The off-axis angle (for superbeams).
\item
 Potential hybrids of different experiments.
\end{enumerate}
Whereas the baseline and parent energy can be easily optimized for, one can, for example, only optimize for
systematical errors or detector properties in this framework by identifying the quantities critical to the physics output, which is of interest for the  experimentalists to focus their resources.

\begin{figure*}[t!]
\begin{center}
\includegraphics[width=0.94\columnwidth]{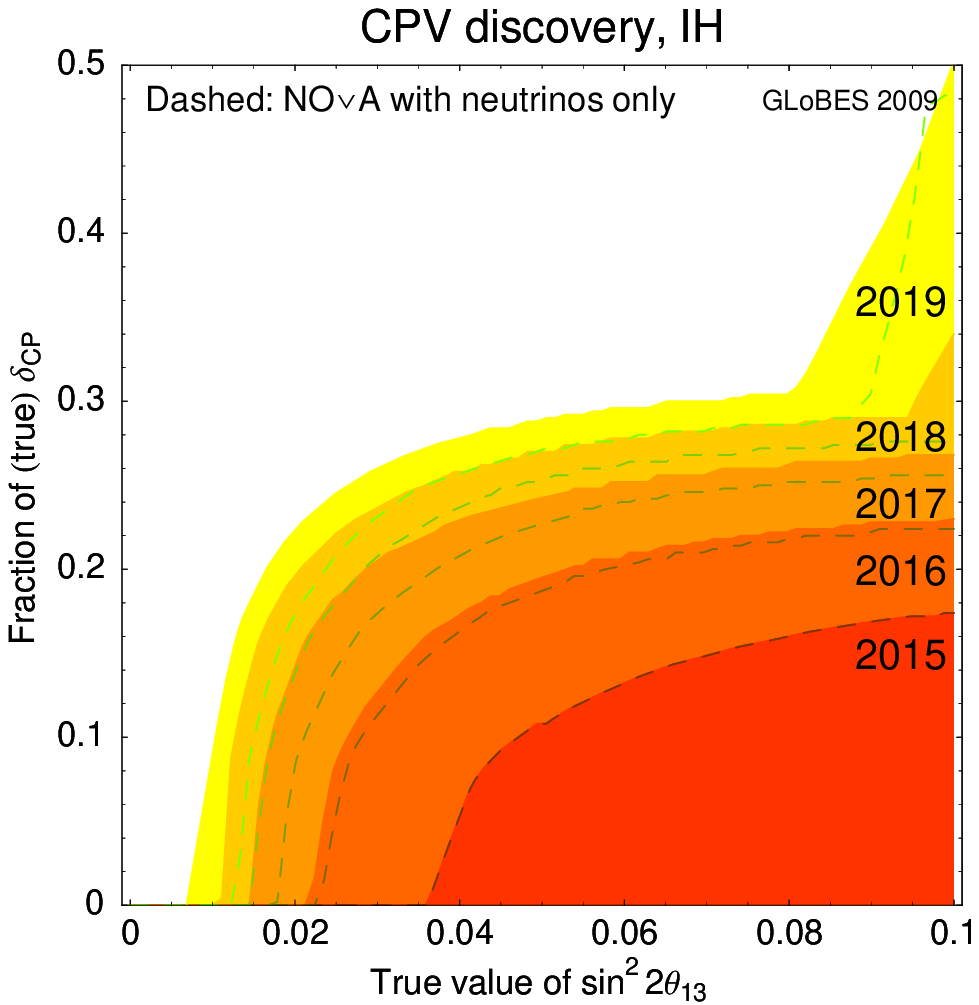} \raisebox{-0.3cm}{
\includegraphics[width=\columnwidth]{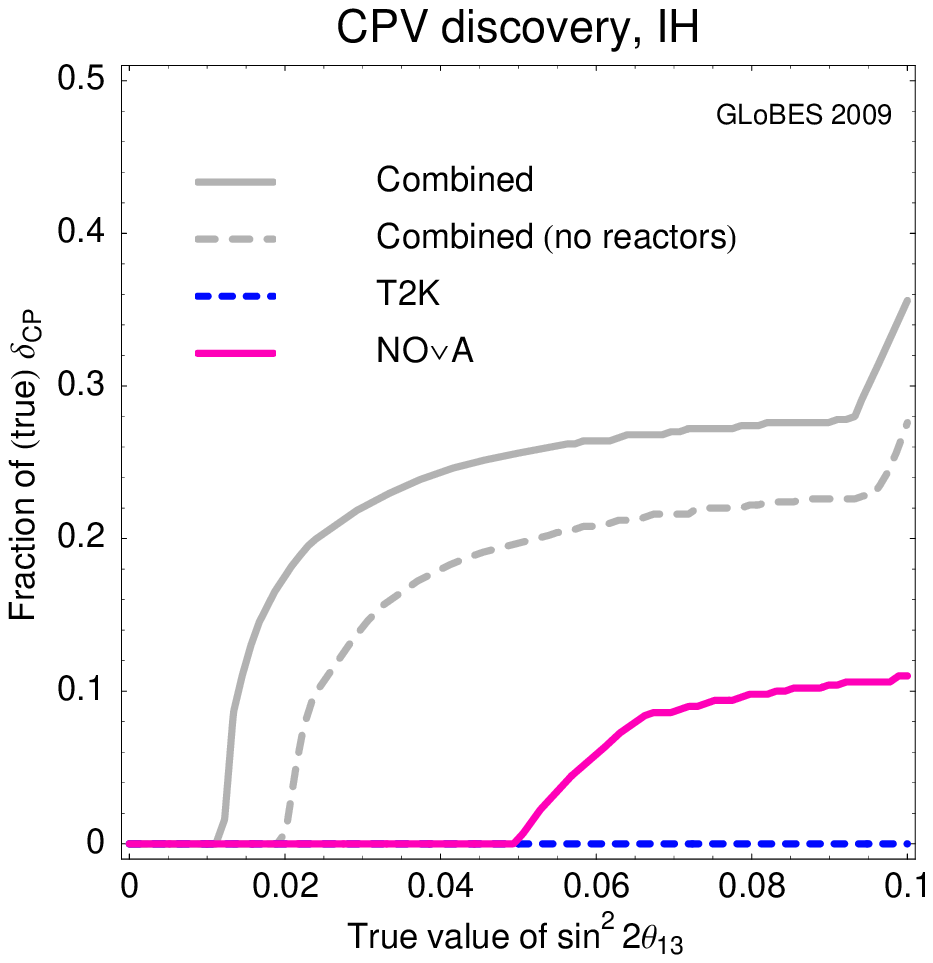}}

\vspace*{-1cm}
\end{center}
\caption{\label{fig:cpv}CP violation discovery reach (90\% CL) simulated for the T2K, NO$\nu$A, Double Chooz, and Daya Bay experiments as a function of the year (left panel) and for different experiments separately (right panel). Inverted mass ordering assumed. Figure from \Ref~\cite{Huber:2009cw}.}
\end{figure*}

For the simulations, often the publically available \GLOBES\ (``General Long Baseline Experiment Simulator'') software~\cite{Huber:2007ji,Huber:2004ka} is used. This is a  multi-purpose software for the simulation of individual long-baseline and reactor neutrino oscillation experiments, as well as for the global analysis of multiple experiments. It includes the treatment of statistics, systematics, correlations, and degeneracies. It consists of two major components:
 Abstract Experiment Definition Language (\AEDL ) describes individual experiments using plain text files, and 
a user interface (C library)  for the $\chi^2$ analysis, which loads one or more \AEDL\ files and provides the functionality for the statistical analysis. 
The separation between \AEDL\ and the user interface makes \GLOBES\ an interesting tool for both the experimentalist and theorist. For example, the theorist may use pre-defined files for the simulation of new, potentially interesting physics effects. The experimentalist, on the other hand, can quickly test the impact of modifications in the experiment definition on physics. Note that \GLOBES\ is not meant to replace a full Monte Carlo simulation of the experiment, but has to be understood as a tool to identify the key parameters and critical factors for especially future experiments. For example, the detector is simulated by an effective response function.
This response function can be used from Monte Carlo simulations as an input for \GLOBES .

\AEDL\ describes an experiment, such as by source type and spectrum, matter density profile, cross sections, detector properties (efficiencies, energy resolution, backgrounds), and systematics. It uses three building blocks: A {\em channel} links a produced flavor state with a certain flux, via the oscillation physics, to the detection with a specific interaction type and the respective cross sections. It results in the event rate of this interaction type. A {\em rule} combines the event rates from different channels, which can either be signal or background for that rule, with a specific systematics; it results in a $\chi^2$. An {\em experiment} contains one or more rules, which are combined to the total $\chi^2$. It shares certain characteristics among the rules, such as baseline and matter density profile, but not the systematical errors.  For example, a simple neutrino factory may store $\mu^+$, which leads to $\nu_e$ and $\bar{\nu}_\mu$ in the beam. A signal channel might be $\nu_e \rightarrow \nu_\mu$, which can be combined into an appearance rule with the background channel $\bar{\nu}_\mu \rightarrow \bar{\nu}_\mu$ (for the charge mis-identified events) leading to a $\chi^2$. An experiment may contain more such rules, such as for different appearance and disappearance channels and different polarities of the initial muons.

One or more descriptions of experiments can be loaded by the C user interface. This interface provides the functionality to extract physical information from the simulated event rate spectra. For example, it allows for sections and projections (marginalizations) of the multi-dimensional fit manifold, \ie, it allows for the inclusion of correlations and degeneracies. Of course, one can also obtain low-level information, such as oscillation probabilities and event rates.
New features in \GLOBES~3.0 are the fully customizable systematics interface supporting multiple sources and detectors (such as for reactor experiments), and the fully customizable external input to be added to the $\chi^2$ before marginalization. 
Heart of the $\chi^2$ analysis is the oscillation and rate engine, which includes a full three-flavor treatment, the use of arbitrary matter density profiles, and an extremely high numerical efficiency with specifically designed numerical algorithms, such as \Ref~\cite{Kopp:2006wp}. The oscillation engine can be modified as well, which allows for the simulation of new physics effects.

The results obtained with the \GLOBES\ software can nowadays be found in the core of many long baseline experiment studies and major collaborative efforts, such as the international neutrino factory and superbeam scoping study (ISS)~\cite{Bandyopadhyay:2007kx} and the US long baseline neutrino experiment study~\cite{Barger:2007yw}, among many others. In addition, application software, such as for multi-parameter marginalization, is available~\cite{Blennow:2009pk}.

An example for such a simulation projecting the CP violation discovery reach is shown in \figu{cpv} (for details, see \Ref~\cite{Huber:2009cw}). The exact sensitivity will depend on a number of factors, such as individual experiment operation plans and data releases. However, such a simulation may give an idea of the information expected at a certain time. For example, one can read off from \figu{cpv} that at the 90\%CL, CP violation will be discovered for at most 50\% of all values of $\deltacp$ until about 2019. Given the low confidence level and the low parameter space coverage, a new generation of experiments may be necessary for a high confidence CP violation discovery. Comparing \figu{cpv} to \figu{cpdid} (lower panel) for a neutrino factory, the latter could be such an instrument. 

\subsection{New physics searches}

Apart from standard oscillation measurements, future neutrino oscillation facilities will be used for new physics searches. Here we show a few possible effects to be tested at the example of a neutrino factory, especially with the help of near detectors.

If the new physics comes from heavy mediators integrated out at low energies, which applies to TeV- or GUT-scale physics, the  effects can be described by a tower of effective operators using the SM fields~\cite{Weinberg:1979sa,Wilczek:1979hc,Buchmuller:1985jz}
\begin{equation}\label{equ:leff}
\mathscr{L} = \mathscr{L}_{\rm SM} + \mathscr{L}^{d=5}_{\text{eff}}
+ \mathscr{L}^{d=6}_{\text{eff}} + \cdots
\end{equation}
with
\begin{equation}
\mathscr{L}^{d}_{\text{eff}} \propto \frac{1}{\Lambda^{d-4}} \, \mathcal{O}^{d}\, ,
\end{equation}
which are invariant under the SM gauge group. Here $\Lambda$ is the new physics scale.
The lowest order addition to the SM is the Weinberg operator in \equ{weinberg}, leading
to Majorana neutrino masses. Therefore, it may be the next logical step to
discuss the implications of the higher dimensional operators. Assuming LHC-scale
physics at $\Lambda \sim 1 \, \mathrm{TeV}$, one can generically estimate that the $d=6$ operators
are suppressed by $(100 \, \mathrm{GeV}/1 \, \mathrm{TeV})^2 \simeq 0.01$ and that the $d=8$ operators
are suppressed by $(100 \, \mathrm{GeV}/1 \, \mathrm{TeV})^4 \simeq 0.0001$ compared to the SM.
In neutrino oscillations, typically percent level effects may be observable, \ie, the effects from $d=6$ operators, whereas the effects from higher dimensional operators will be very hard to access. Therefore, we focus on $d=6$ operators in the following which are generated at tree-level.

The first class of effective $d=6$ operators of interest are so-called \hl{non-standard interactions} (NSI)
\begin{equation}
\mathscr{L}^{d=6}_{\mathrm{NSI}} = 2\, \sqrt{2} \, G_F \, \varepsilon^{\alpha
\gamma}_{\beta \delta} \,
 \left(
 \bar{\nu}^{\beta} \gamma^{\rho} {\rm P}_{L} \nu_{\alpha}
  \right) \,
 \left(
 \bar{\ell}^{\delta} \gamma^{\rho} {\rm P}_{L/R} \ell_{\gamma}
 \right) \, ,
\label{equ:nsi}
\end{equation}
where $\ell$ denote the charged leptons. Here $G_F$ is the Fermi
coupling constant and $P_L$ and $P_R$ are the left- and right-handed
(chiral) projection operators, respectively. Such operators
lead to \hl{NSI matter effects} adding to the Hamiltonian in
\equ{ham3} (for $\gamma=\delta=e$ in \equ{nsi}):
\begin{equation}
\delta \mathcal{H}_F = V_{\mathrm{CC}} \, \left( 
\begin{array}{ccc}
\varepsilon_{ee}^m & \varepsilon_{e \mu}^m & \varepsilon_{e \tau}^m  \\
(\varepsilon_{e \mu}^m)^*  & \varepsilon_{\mu \mu}^m  & \varepsilon_{\mu \tau}^m  \\
(\varepsilon_{e \tau }^m)^* & (\varepsilon_{\mu \tau}^m)^* & \varepsilon_{\tau \tau }^m 
\end{array}
\right) \, .
\end{equation}
Note that $\varepsilon$ is the strength of the NSI effect relative to the SM matter effect.
In addition to the propagation in matter, the production or detection
processes can be affected by NSI. The neutrino states produced in a
source and observed at a detector can be treated as superpositions
of pure orthonormal flavor states \cite{Grossman:1995wx,Gonzalez-Garcia:2001mp,Bilenky:1992wv}:
\begin{eqnarray}\label{equ:s}
|\nu^s_\alpha \rangle & = &  |\nu_\alpha \rangle +
\sum_{\beta=e,\mu,\tau} \varepsilon^s_{\alpha\beta}
|\nu_\beta\rangle   \ , \\
\langle \nu^d_\beta| & = &\langle
 \nu_\beta | + \sum_{\alpha=e,\mu,\tau}
\varepsilon^d_{\alpha \beta} \langle  \nu_\alpha  |  \ . \label{equ:d}
\end{eqnarray}
For instance, for neutrino production by muon decays at a neutrino factory, one obtains $\varepsilon^s_{\mu \beta}$ for $\alpha=\delta=e$ and $\gamma=\mu$ in \equ{nsi}. Note that these $\varepsilon^s$ and $\varepsilon^d$ are process dependent quantities.

In writing down \equ{nsi}, we do not require gauge invariance. If
SU(2) gauge invariance is imposed at the effective operator level,
typically charged lepton flavor violating processes will be induced
because the neutrinos come together with charged leptons in SU(2) doublets.
If it is required that all the charged-lepton processes vanish,
only the NSI operators made out of four lepton doublets survive which are antisymmetric in the flavor indices,
\ie, $ \alpha \neq \gamma$ and $\beta \neq \delta$. Such operators
can be naturally realized in theories with an SM SU(2) singlet
singly charged
scalar~\cite{Bergmann:1999pk,Bilenky:1993bt,Cuypers:1996ia}. These models
are, however, strongly constrained otherwise, such as by lepton universality tests~\cite{Antusch:2008tz}.
Therefore, it is difficult to find a model for large NSI from leptonic $d=6$ operators, and for higher dimensional operators a model cannot be easily found without cancellations~\cite{Antusch:2008tz,Gavela:2008ra}. In summary, the model-independent NSI bounds are rather weak~\cite{Biggio:2009nt} and deserve a further test at future experiments. However, the prospects for $d=6$ operators (generated at tree-level) are not very good from a theoretical perspective, which means that large NSI should come from higher dimensional or loop-induced operators. But these are generically expected to be much smaller than the tree-level $d=6$ contributions, quite likely beyond the reach of future experiments unless the new physics scale is very close to the EWSB scale. 

Using neutrino factory near detectors, especially $\varepsilon_{e \tau}^s$ and $\varepsilon_{\mu \tau}^s$ are interesting to be tested, because the neutrino factory beam does not contain tau neutrinos. The expected sensitivity for an OPERA-like detector is at the level of $10^{-3}$ to $10^{-4}$ at the 90\%CL~\cite{Tang:2009na}, maybe at the level where one may expect some $d=8$ contributions.
For matter NSI, the expected sensitivity for the NSI including the tau sector is about $10^{-2}$ at $3 \sigma$ from the long baselines~\cite{Kopp:2008ds}, which is at least an order of magnitude beyond the current model-independent bounds, but maybe too large for suspecting a $d=8$ contribution. Assuming $d=6$ operators without charged lepton flavor violation, certain correlations between source and matter NSI are present~\cite{Gavela:2008ra}, which lead to an enhanced sensitivity~\cite{Tang:2009na}. However, in this case, the sensitivity has to be compared to the model-dependent bounds, which it exceeds only by about a factor of two~\cite{Meloni:2009cg}.

Another class of effective $d=6$ operators are coming from integrating out heavy fermion fields, leading to
 \hl{non-unitarity} (NU) of the PMNS matrix. Such fermions are often introduced in seesaw mechanisms at the TeV scale. In general, gauge invariant theories extending the SM with the
tree-level exchange of heavy neutral fermions result in a
dimension-six operator of the form~\cite{Abada:2007ux,Antusch:2006vwa}
\begin{eqnarray}
\mathscr{L}^{d=6}_{\mathrm{NU}}  = c_{\alpha\beta} \left(\overline{L}_\alpha
\tilde H \right) {\rm i} \slc\partial \left(\tilde H^\dagger
L_\beta\right) \,  \label{equ:LNU}
\end{eqnarray}
with $\tilde H= i \sigma^2 H$. 
Re-diagonalizing and re-normalizing the kinetic terms of the neutrinos, one has an effective Lagrangian
\begin{eqnarray}
\mathscr{L}^{\mathrm{eff}} & = & \frac{1}{2} ( \bar \nu_i i \slc\partial \nu_i - \overline{\nu^c}_i m_i \nu_i ) \nonumber \\
& - & \frac{g}{2 \sqrt{2}} (W^+_\mu \bar{\ell}_\alpha \gamma_\mu (1 - \gamma_5) N_{\alpha i} \nu_i ) \nonumber \\ 
& - & \frac{g}{2 \cos \theta_W} (Z_\mu \bar\nu_i \gamma^\mu (1-\gamma_5) (N^\dagger N)_{ij} \nu_j) \nonumber \\
& + & \mathrm{H.c.}
\end{eqnarray}
with modified couplings to the $W$ and $Z$ bosons. Here $N$ is an effective (non-unitary) mixing matrix which can be parameterized by 
\begin{equation}
N=(1+\varepsilon) U \, .
\end{equation}
Because neutrino oscillations are typically tested at energies below the gauge boson masses where the gauge bosons are effectively integrated out, the modified couplings effectively lead to non-standard four fermion interactions of the type \equ{nsi}. However, the source, detector, and matter NSI are correlated in a particular, fundamental way (\ie, process-independent), which leads to an enhanced sensitivity. Especially the tau sector benefits from a near detector~\cite{Antusch:2009pm}. In fact, at the neutrino factory NSI from $d=6$ effective operators and NU lead to very similar effects for $\epsilon_{\mu \tau}$, because the correlation between source and matter NSI is basically the same~\cite{Meloni:2009cg}. However, note that the source NSI are process-dependent, whereas the source NU are fundamental, which means that with the help of a superbeam-based experiment the effects could be disentangled, at least in principle.

\begin{figure}[t]
\begin{center}
\includegraphics[width=\columnwidth]{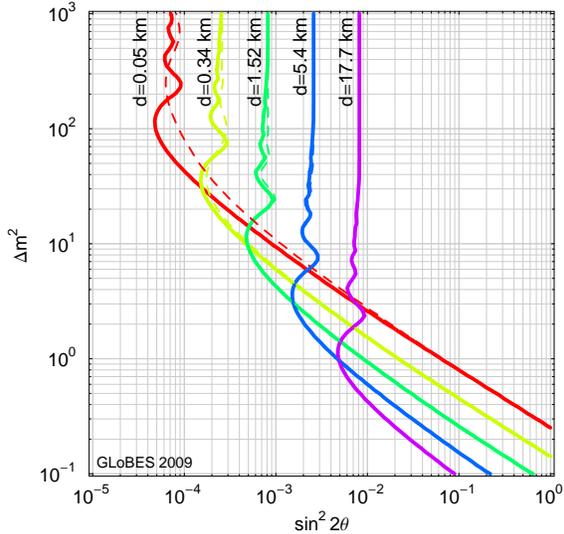}
\end{center}
\vspace*{-1cm}
\caption{\label{fig:es} Exclusion limit for sterile neutrinos measured in the two-flavor limit of $P_{ee}$ for several near detector distances $d$ (distance to the end of the decays straight) and ideal, systematics-free near detectors at a neutrino factory (90\% CL, 2 d.o.f.; two near detectors in front of straights). The dashed curves illustrate the effect of including the averaging over the decay straight,
whereas the solid curves are without this averaging. The fiducial detector masses are fixed to 200~kg. Note that there is no systematics included in this figure. Taken from \Ref~\cite{Giunti:2009en}.}
\end{figure}

If the neutral fermion fields are light enough to be produced in the neutrino oscillation experiment, they enter \equ{mix} directly as sterile (not weakly interacting) states. Although one does not expect a major contribution of these sterile states anymore, such as to describe the LSND anomaly~\cite{Aguilar:2001ty}, small ad-mixtures of sterile neutrinos are not ruled out and are a clear signature for new physics. If the sterile neutrino mass splitting is significantly above $\ldm$, \ie, $\Delta_{41} \gg \Delta_{31} \gg \Delta_{21}$, sterile neutrinos are best searched for at short baselines where $\Delta_{31} \sim \Delta_{21} \sim 0$. Therefore, a prominent model-independent way to search for sterile neutrinos is testing \equ{twoflavor} in various oscillation channels at baselines where standard oscillations cannot have developed yet. A number of experiments have done such tests in the past, such as NOMAD~\cite{Altegoer:1997gv} and CHORUS~\cite{Eskut:1997ar} including $\nu_\tau$ appearance.
Similar tests can be performed at a neutrino factory.
For example, take the electron neutrino disappearance probability
\begin{equation}
P_{ee}  =  1 - \sin^2  2 \theta  \,  \sin^2 \left( \frac{ \Delta m^2 L}{4 E} \right) \, .
\label{equ:ee}
\end{equation}
Then one typically shows exclusion limits of the form in \figu{es}: For each curve, the r.h.s. is excluded at the given confidence level. In this case, $\theta=0$ is simulated in the data, and the $\theta$ and $\Delta m^2$ in the figure correspond to the fit values. Obviously, the optimum sensitivity (peak) depends on the choice of the baseline $L$, related to $d$ in this case. Here $d$ is the distance to the end of the decay straight: since the neutrino factory is a line source, the baseline $L$ is not defined for short distances. The effect of the averaging over the line source is shown as dashed curves.  The sensitivity typically breaks away for small $\Delta m^2$, because for too small arguments of the sine in \equ{ee} the oscillation does not develop and $\theta$ cannot be measured. Maximal sensitivity in the $\theta$ direction is obtained at the first oscillation maximum $\Delta = \Delta m^2 L/(4 E) \simeq \pi/2$, where the spectral effect can be easiest measured. Then higher oscillation maxima are visible, until the oscillation averages out. For large $\Delta m^2$, typically the total event rate or systematics limits the sensitivity. The dependence of the sensitivity on the baseline is characteristic for the sterile neutrino example, whereas for the NSI and NU near detection the baseline choice only affects statistics. Note that such figures are not only used for new physics searches, but also for the $\ldm$-$\theta_{13}$-exclusion region; see, \eg, Fig.~3 in \Ref~\cite{Schwetz:2008er}.

In summary, future facilities can be used to constrain a number of new physics effects, where we have only shown some examples here. Whereas the physics from higher dimensional operators may be most interesting in the context of LHC physics, the sterile neutrino example illustrates that also the location of a near detector system of future experiments may be important, and needs to be taken into account in the optimization.

\section{Neutrino oscillations in the Sun}
\label{sec:sun}

Neutrino astronomy is an emerging field of neutrino physics, which is so far based on the observation of solar and supernova neutrinos. In 2002, Raymond Davis Jr and Masatoshi Koshiba received the Nobel prize in physics ``for pioneering contributions to astrophysics, in particular for the detection of cosmic neutrinos''. On the one hand,  Raymond Davis Jr and his collaboration detected over three decades 2000 neutrinos from the Sun, which is an important piece of evidence for nuclear fusion in the Sun‘s interior. On the other hand, Masatoshi Koshiba and collaborators detected on February 23th, 1987 twelve of the  $10^{16}$ neutrinos which passed their detector  from an extragalactic supernova explosion. Both of these observations can be regarded as the foundation of  neutrino astronomy.

Especially neutrino oscillations in the Sun have also contributed to the measurement of the neutrino properties. Nowadays they still provide the most accurate measurement of the solar mixing angle $\theta_{12}$. Neutrino oscillations in the Sun and in supernovae can, however, not be treated within the framework of the previous section, because the matter density is varying along the propagation path and not constant. In order to illustrate the differences to constant matter, we focus on the two-flavor case in the Sun in this section, and show how it has lead to the measurement of $\theta_{12}$.  The description of neutrino oscillations in supernovae is more complicated, because the high neutrino densities lead to neutrino self interactions, which again lead to collective phenomena (see, \eg, \Refs~\cite{Raffelt:2007cb,Dasgupta:2007ws}). Note that supernova neutrinos might also be used for the determination of neutrino properties, such as the observation of SN1987A has already lead to a bound for neutrino lifetime. However, for neutrino masses and mixings, the predictions strongly depend on the parameters of the source and the availability of detectors.

Neutrinos are assumed to be produced in the deep interior of the Sun, oscillate within the Sun until they reach its surface, and then propagate as mass eigenstates to the Earth. Therefore, there are no neutrino oscillations between Sun and Earth in our current understanding. As we shall see below, this can be naturally understood in terms of mass eigenstates emitted from the Sun. However, even if there was a superposition of states emitted, coherence between Sun and Earth would be eventually lost, and mass eigenstates arrived at the Earth's surface. Therefore, we only deal with neutrino oscillations within the Sun in this section. Note, however, that the neutrinos start to oscillate again once they enter Earth matter. Therefore, if they pass substantial Earth matter before they are detected, \ie, they are detected on the ``night'' side of the Earth coming from below, some oscillating effect may be visible. The difference between direct detection and detection after the propagation in Earth matter is also called \hl{day-night-effect}. This effect has not been observed yet, which is not surprising: Applying the solar parameters to \equ{eres}, we obtain a resonance energy of a few hundred MeV. Solar neutrinos only extend up to about 18~MeV, which is far below this resonance. Therefore, only small effects can be expected. In supernova neutrinos, however, there is a tail of neutrinos extending to much higher energies $\mathcal{O}(100 \, \mathrm{MeV})$, where the Earth matter effects may be visible (see, \eg, \Refs~\cite{Lindner:2002wm,Dighe:2003jg} for different applications).

Let us now first recall the differences between neutrino oscillations in vacuum and matter. 
In vacuum, we have -- \cf, \equ{hvac}
\begin{eqnarray}
\mathcal{H}& =& U \mathcal{H}_{\mathrm{diag}} U^\dagger   \\
& \rightarrow& \frac{1}{4E} \left( 
\begin{array}{cc}
-\Delta m_{21}^2 \cos 2 \theta_{12}  & \Delta m_{21}^2 \sin 2 \theta_{12} \\
 \Delta m_{21}^2 \sin 2 \theta_{12} & \Delta m_{21}^2 \cos 2 \theta_{12} 
\end{array}
\right) \, ,\nonumber
\end{eqnarray}
where the arrow refers to the subtraction of an overall phase factor.
Here 
\begin{eqnarray}
U & = &  \left(
\begin{array}{cc}
\cos \theta_{12} & \sin \theta_{12} \\
-\sin \theta_{12} & \cos \theta_{12}
\end{array}
\right) \, , \nonumber \\
\mathcal{H}_{\mathrm{diag}} & = & \frac{1}{2 E} \left(
\begin{array}{cc}
0 & 0 \\
0 & \Delta m_{21}^2
\end{array}
\right) \, .
\end{eqnarray}
Similarly -- \cf, \equ{hmatter2} -- we have in matter
\begin{eqnarray}
\mathcal{\tilde H} & = &
\tilde U 
\mathcal{\tilde H}_{\mathrm{diag}} 
\tilde U^\dagger  \\
& \rightarrow & \frac{1}{4E} \left( 
\begin{array}{cc}
-\Delta \tilde m^2 \cos 2 \tilde \theta  & \Delta \tilde m^2 \sin 2 \tilde \theta \\
 \Delta \tilde m^2 \sin 2 \tilde \theta & \Delta \tilde m^2 \cos 2 \tilde \theta 
\end{array}
\right) \nonumber
\end{eqnarray}
with
\begin{eqnarray}
\tilde U & = &\left(
\begin{array}{ cc}
\cos \tilde \theta & \sin \tilde \theta \\
-\sin \tilde \theta & \cos \tilde \theta
\end{array}
\right)  \nonumber \\ 
\mathcal{\tilde H}_{\mathrm{diag}} & = & \frac{1}{2 E} \left(
\begin{array}{cc}
0 & 0 \\
0 & \Delta \tilde  m^2
\end{array}
\right) \, ,
\end{eqnarray}
using the parameter mapping in \equ{mapping}. In this case, the eigenstates of the Hamiltonian and the flavor eigenstates are connected similar to \equ{mix}, with  $U$ replaced by $\tilde U$. Therefore, one refers to the eigenstates of the Hamiltonian as \hl{mass eigenstates in matter}. In constant matter density, the states can be propagated using the evolution operator
\begin{eqnarray}
e^{-i \mathcal{\tilde H} L} = \tilde{U} e^{-i \mathcal{\tilde H}_{\mathrm{diag}} \, L} \tilde{U}^\dagger  ,
\end{eqnarray}
because the Hamiltonian is not explicitely time-dependent. In varying matter density, however, the full Schr{\"o}dinger equation has to be used
\begin{eqnarray}
\label{equ:sg}
& & i \frac{d}{dx} \left( 
\begin{array}{c}
\psi_e \\ \psi_x
\end{array} \right)
 =  \\ & & \quad \frac{1}{4 E}
\left(
\begin{array}{cc}
- \Delta \tilde m^2 \cos 2 \tilde \theta & \Delta \tilde m^2 \sin 2 \tilde \theta \\
\Delta \tilde m^2 \sin 2 \tilde \theta &  \Delta \tilde m^2 \cos 2 \tilde \theta
\end{array}
\right)
\left( 
\begin{array}{c}
\psi_e \\ \psi_x
\end{array} \right) \, ,\nonumber
\end{eqnarray}
where $\psi_e$ and $\psi_x$ are the amplitudes for being $\nu_e$ or $\nu_x$. Here $\nu_x$ is a superposition of $\nu_\mu$ and $\nu_\tau$, because these are maximally mixed. For example, in the beginning, we have electron neutrinos, and therefore $\psi_e = \langle \nu_e | \nu_\mathrm{in} \rangle = 1$ and $\psi_x = \langle \nu_x | \nu_\mathrm{in} \rangle = 0$. As the next step, one uses a transformation to the mass eigenstates in matter $\phi_i$
\begin{equation}
\left( \begin{array}{c}
\psi_e \\ \psi_x
\end{array} \right)
=
\tilde U \,  \left( \begin{array}{c}
\phi_1 \\ \phi_2
\end{array} \right) \, .
\end{equation}
Applying this transformation to \equ{sg}, one finds
\begin{equation}
i \frac{d}{dx} \left( 
\begin{array}{c}
\phi_1 \\
\phi_2
\end{array}\right) 
=
\frac{1}{4E}
\left(
\begin{array}{cc}
- \Delta \tilde m^2 & -4 E i \frac{\partial \tilde \theta}{\partial x} \\
4 E i \frac{\partial \tilde \theta}{\partial x} & \Delta \tilde m^2 
\end{array}
\right)
\left( 
\begin{array}{c}
\phi_1 \\
\phi_2
\end{array}\right) \, , 
\end{equation}
which is a coupled differential equation system.  One can easily see that the 
differential equations decouple if the diagonal entries in the matrix are
dominant, \ie, 
\begin{equation}
|\Delta \tilde m^2| \gg 4 E \left| \frac{\partial \tilde \theta}{\partial x} \right| \, .
\end{equation}
This electron density profile-dependent condition (or derivations from it) is also called the \hl{adiabaticity condition}, and neutrino oscillations
in the Sun can be described in the \hl{adiabatic limit} to a good approximation. Using this condition,
each of the differential equations can be easily solved in order to obtain $\phi_i(x) = \exp(i \xi_i(x)) \, \phi_i(0)$ with a phase factor $\xi_i(x)$ depending on the matter density profile.

\begin{figure}[t!]
\begin{center}
\includegraphics[width=\columnwidth]{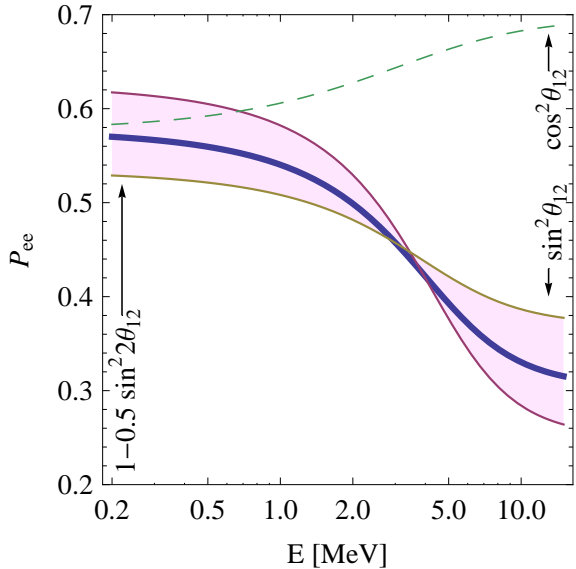}

\vspace*{-1cm}
\end{center}
\caption{\label{fig:sun} Electron neutrino disappearance probability for solar neutrinos as a function of energy in the perfectly adiabatic limit. The thick curve represents the current best-fit values of the solar parameters, the thin curves limit the $3\sigma$ allowed range for $\sin^2 \theta_{12}$ (\cf, \Tab~\ref{tab:summary}). The dashed curve shows the solution for the inverted mass ordering.}
\end{figure}

Let us now discuss the simplest case $\hat{A} \gg 1$ in \equ{ahat} at the production point. That implies both high enough densities at the production point and large enough neutrino energies. Then we can read off from 
\equ{mapping} that $\sin 2 \tilde \theta \rightarrow 0$ or $\tilde \theta = \pi/2$ (which is the solution for $\Delta m_{21}^2>0$). At the production point $x=0$, where the neutrinos are produced as electron neutrinos, we therefore have
\begin{eqnarray}
& & \psi(x=0)  =  \left( \begin{array}{c} 1 \\ 0 \end{array} \right)  \nonumber \\
& \Rightarrow &
\phi(x=0) = \tilde U^{-1} \left( \begin{array}{c} 1 \\ 0 \end{array} \right) \nonumber \\
& = &
 \left( \begin{array}{c} \cos \tilde \theta \\ \sin \tilde \theta \end{array} \right) \overset{\hat A \gg 1}{\longrightarrow}   \left( \begin{array}{c} 0  \\ 1 \end{array} \right)   \, .
\label{equ:final}
\end{eqnarray}
At an arbitrary point $x$ after propagation, we find
\begin{eqnarray}
& & \left( \begin{array}{c} \psi_e \\ \psi_x \end{array} \right) (x) =  \tilde U(x) 
\left( \begin{array}{c} \phi_1 \\ \phi_2 \end{array} \right) (x)  \\
& & =    \tilde U(x) 
\left( \begin{array}{c}  e^{i \xi_1(x)} \, \phi_1(0) \\  e^{i \xi_2(x)} \, \phi_2(0) \end{array} \right)  =   \tilde U(x) 
\left( \begin{array}{c} 0 \\ e^{i \xi_2(x)} \end{array} \right)  \, . \nonumber
\end{eqnarray}
In the Sun, the electron density drops approximately exponentially as
\begin{equation}
n_e(x) = n_e(0) \exp \left( - \frac{x}{r_0} \right) \, , \quad r_0 \simeq \frac{R_\odot}{10} \, .
\end{equation}
This means that for $x \rightarrow \infty$, the electron density drops continuously to the vacuum density, and we have
\begin{eqnarray}
& & \left( \begin{array}{c} \psi_e \\ \psi_x \end{array} \right) (x \rightarrow \infty)  \overset{\tilde U \rightarrow U}{=}    U 
\left( \begin{array}{c} 0 \\ e^{i \xi_2(x)} \end{array} \right)  \nonumber \\
& & \qquad = \left( \begin{array}{c} \sin \theta_{12} \\ \cos \theta_{12} \end{array} \right) e^{i \xi_2(x)} 
\end{eqnarray}
Finally, we obtain for the electron neutrino disappearance probability
\begin{equation}
P_{ee} = \left|   \left( 
\begin{array}{cc}
1 & 0
\end{array} 
\right)
\left(
\begin{array}{c}
\psi_e \\ \psi_x
\end{array}
\right)
(t \rightarrow \infty) 
\right|^2
= \sin^2 \theta_{12} \, ,
\label{equ:solad}
\end{equation}
\ie, the phase factor, which we have not computed explicitely, drops out. 

The factor $\sin^2 \theta_{12}$ describes the disappearance of electron neutrinos from the Sun through neutrino oscillations in the Sun's interior at high enough energies, such as measured by the SNO experiment. In practice, the condition $\hat A \gg 1$ used for this derivation only holds for $E \gtrsim 10 \, \mathrm{MeV}$, whereas for very small energies $E \lesssim 1 \, \mathrm{MeV}$ one basically obtains the vacuum result ($\hat A \simeq 0$) with averaged oscillations (\cf, \equ{average} applied to \equ{twoflavor})
\begin{equation}
P_{ee} = 1 - \frac{1}{2} \sin^2 2 \theta_{12} \, .
\label{equ:solvac}
\end{equation}
For the intermediate case, one finds a continuous transition depending on the size of $\hat{A}$. For $\Delta m_{21}^2 >0$, the effective mixing angle $\tilde \theta$ in \equ{mapping} starts at the vacuum value $\theta$ at low energies, increases to $\pi/4$ at the resonance energy (where $\sin2 \tilde \theta=1$), flips the octant and increases further to $\pi/2$. For $\Delta m_{21}^2 < 0$, there is no resonance, and the angle decreases continuously from the vacuum angle $\theta$ to $0$. In this case, $\sin 2 \tilde \theta \rightarrow 0$ at high energies evaluates to $\tilde \theta \rightarrow 0$. One obtains $\phi(x=0)=(1,0)^T$ in \equ{final}, leading to $P_{ee} = \cos^2 \theta_{12}$. We illustrate the transition probability as a function of energy in \figu{sun} for the perfectly adiabatic case and the parameters from \Tab~\ref{tab:summary} (thick curve). In this figure, the lower energy limit corresponds to \equ{solvac}, the upper energy limit end to \equ{solad}. The thin curves limit the $3 \sigma$ allowed range for $\sin^2 \theta_{12}$ (\cf, \Tab~\ref{tab:summary}). Note that there is little dependence on $\sdm$ in its $3\sigma$ allowed range, because $\sdm$ is very well measured by the KamLAND experiment. The dashed curve shows the $\Delta m_{21}^2 < 0$ case.
The energy dependence in \figu{sun} has been measured by early Gallium and the Homestake experiments at the very left end, and by SNO at the very right end. Therefore, the $\Delta m_{21}^2 < 0$ case has been excluded from the solar neutrino observations.\footnote{Strictly speaking, this discussion can only be done together with the choice of the octant of $\theta_{12}$, since there is an ambiguity $\sin^2 \theta_{12} \rightarrow \cos^2 \theta_{12}$, $\Delta m_{21}^2 \rightarrow - \Delta m_{21}^2$ in these derivations, \ie, instead of changing the sign of $\Delta m_{21}^2$, one can also change the octant of $\theta_{12}$. In either case, there are two qualitatively different cases (resonance/no resonance), which can be distinguished.} In the future, the BOREXINO experiment~\cite{Alimonti:2000xc} has the potential to improve the information in the intermediate energy range, and verify the energy dependence which is so characteristic of the solar flavor transitions.
Note that the flavor transitions described in this section are also often called the \hl{MSW effect}, named after Mikheyev, Smirnov, and Wolfenstein~\cite{Mikheev:1985gs,Mikheev:1986wj,Wolfenstein:1978ue}. The case of constant matter density in the previous section is a special case of the general MSW effect.

\section{Neutrinos from cosmic accelerators}
\label{sec:cosm}

High energetic neutrinos are especially produced in the Earth's atmosphere or in man-made terrestrial experiments. However, above the TeV-boundary, also \hl{cosmic accelerators} may produce neutrinos, see \Refs~\cite{Learned:2000sw,Becker:2007sv,Chiarusi:2009ng} for review articles. In particular, the observation of high energetic cosmic rays, which are believed to come from extragalactic sources at very high energies, motivates the existence of such accelerators. If, however, these hadrons interact with other hadrons or photons in the source, a significant neutrino flux will be produced as well. Experiments such as ANTARES~\cite{Aslanides:1999vq} in the Mediterranean or IceCube~\cite{Ahrens:2002dv} at the South Pole are built for the observation of such fluxes, they are often called \hl{neutrino telescopes}. Known candidates for extragalactic accelerators as potential neutrino sources include \hl{active galactic nuclei} (AGNs)~\cite{Mannheim:1993,Mucke:2000rn,Aharonian:2002} and \hl{gamma ray bursts} (GRBs)~\cite{Waxman:1997ti}, see also \Ref~\cite{Rachen:1998fd} for theoretical considerations. If these sources accelerate enough hadrons, neutrino fluxes are to be expected, which should be observable in the neutrino telescopes.  However, such cosmic neutrinos have not been observed yet, a fact which may be not so surprising from the current point of view. If one relates the possible neutrino flux from such cosmic accelerators to the cosmic ray flux, one obtains an upper bound on the neutrino flux, the so-called Waxman-Bahcall bound~\cite{Waxman:1998yy}. This bound assumes that the neutrons, produced in photohadronic processes, escape from the source before they decay. A different version is the Mannheim-Protheroe-Rachen bound~\cite{Mannheim:1998wp}, which includes sources optically thick to neutrons (the neutrons interact before they can escape), and uses gamma rays as an additional information source. The IceCube experiment will exceed these bounds in the coming years, which means that the detection of cosmic neutrinos in the near future might be quite plausible.

The observation of extragalactic neutrino fluxes may be interesting for different reasons. First of all, the fluxes are evidence for the hadron content in the source. Second, neutrinos directly point to the source, unlike the cosmic rays, which are affected by magnetic fields, and  photons, which are easily absorbed or scattered. And third, such neutrino fluxes may be used for the determination of new physics properties. For example, neutrino properties such as neutrino lifetime will be tested. In this section, we will argue from the source via propagation to detection, focusing on the determination of neutrino properties via propagation effects.

\subsection{Neutrino production in the source}

\begin{figure}[t]
\begin{center}
\includegraphics[width=\columnwidth]{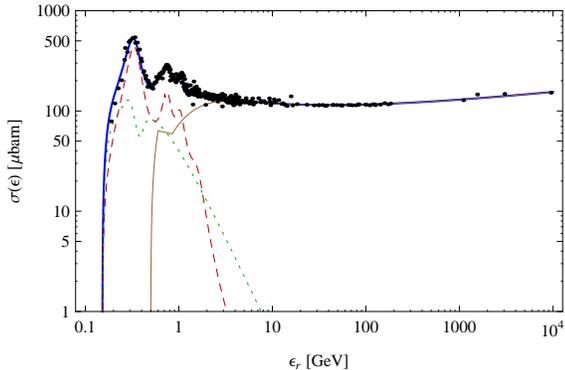}

\vspace*{-0.5cm}

\caption{\label{fig:xsec} The total $p\gamma$ photo-meson cross section 
as a function of the photon energy in the proton rest frame $\epsilon_r$
analog to \cite{Mucke:1999yb} ($1\mu$barn = $10^{-30}$ cm$^{2}$; data, shown as dots, from \Ref~\cite{Amsler:2008zzb})). The contributions of baryon
resonances (red, dashed), the direct channel (green, dotted)
and multi-pion production (brown) are shown separately. Figure taken from \Ref~\cite{Hummer:2010vx}.}
\end{center}
\end{figure}

Neutrinos are typically assumed to be produced by $pp$ interactions, or $p\gamma$ \hl{photohadronic} processes. These interactions lead to charged pions, among other particles, which decay into neutrinos through \equ{pion}. The interacting protons are assumed to be accelerated in relativistic jets by Fermi shock acceleration, which leads to a power law proton injection spectrum. The target protons in $pp$ interactions are typically introduced using external material, such as dust, hit by the relativistic outflow, which needs to be described by additional parameters. In self-consistent models, photohadronic interactions with target photons are used, which originate  from synchrotron radiation of electrons or positrons co-accelerated with the protons. The synchrotron photon field can, to a first approximation, be computed from the model parameters, such as the magnetic field and the spectral index of the electrons (positrons), see, \eg, \Ref~\cite{Reynoso:2008gs} for such an approach. Alternatively, a thermal target photon field, such as from an accretion disk, may be used as a target, which again introduces new parameters.

The photohadronic neutrino production is often described by the $\Delta$-resonance approximation
\begin{equation}
p + \gamma \rightarrow \Delta^+ \rightarrow \left\{
\begin{array}{ll}
n + \pi^+ & 1/3 \, \, \text{of all cases} \\
p + \pi^0 & 2/3 \, \, \text{of all cases} \\
\end{array}
\right. .
\label{equ:ds}
\end{equation}
This description, is however, not sufficient for our purposes. First of all, other processes contribute, as shown in \figu{xsec}, such as direct (t-channel) production or higher resonances. These other processes have different characteristics, such as different energies where the pions are found as a function of the initial proton energy (different inelasticities), and different multiplicities of the pions. 
For example, $\pi^-$ are produced in higher resonances in addition to the $\pi^+$ and $\pi^0$ in \equ{ds}. These $\pi^-$ affect the neutrino-antineutrino ratio, which may be used to test the difference between $pp$ and $p\gamma$ neutrino production (in the $pp$ production $\pi^+$ and $\pi^-$ are produced in equal amounts). The Glashow resonance $\bar{\nu}_e + e^- \to W^- \to \hdots$ at about  $6.3 \, \text{PeV}$~\cite{Learned:1994wg,Anchordoqui:2004eb,Bhattacharjee:2005nh} is a detection process suitable for neutrino-antineutrino discrimination. In addition, the high energy tail of the multi-pion production in \figu{xsec} leads to a change of the spectral shape, as we will see below.
More refined descriptions of the neutrino fluxes from photohadronic interactions therefore use the Monte-Carlo simulation of the SOPHIA software~\cite{Mucke:1999yb,Rachen:1996ph}. For time-consuming source simulations, an effective description such as \Ref~\cite{Kelner:2008ke} is an efficient alternative, or a simplified interaction model, such as \Ref~\cite{Hummer:2010vx} if the intermediate particles, such as muons, are to be treated explicitely to include cooling effects.

\begin{figure*}[t]
\begin{center}
\includegraphics[width=0.8\textwidth]{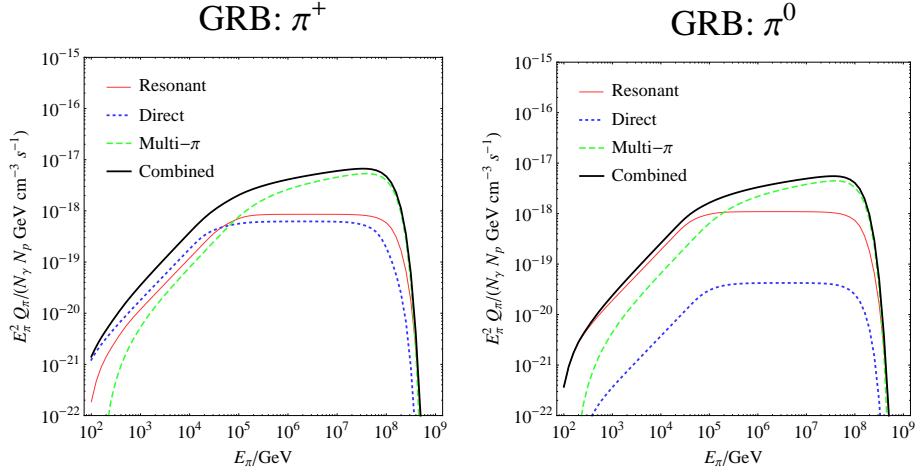}

\vspace*{-0.5cm}

\caption{\label{fig:grb}  Contributions of resonant (thin solid), direct (dotted) and multi-pion (dashed) production for $\pi^+$ (left) and $\pi^0$ (right) spectra for proton-photon interactions using the GRB benchmark from \Ref~\cite{Lipari:2007su}. Figure taken from \Ref~\cite{Hummer:2010vx}, using the description of processes in there.}
\end{center}
\end{figure*}

The individual contributions of the different processes to the $\pi^+$ and $\pi^0$ spectra in a GRB are shown in \figu{grb} in the left and right panels, respectively. In these spectra, the cutoff at high energies is introduced by hand, since cooling processes are not included. The spectrum from resonant production, shown as thin solid curves, resembles the typical GRB spectra often shown in the literature. However, the slope of the plateau is increased by the fact that the cross section in \figu{xsec} is non-vanishing for large energies, as it can be seen by the dashed curves representing the multi-pion contribution, and in fact also depends on the extrapolation to higher energies. For the $\pi^+$ spectrum, the resonant production only dominates in a very narrow region around the first spectral break, whereas the $\pi^0$ spectrum is dominated by the resonant production, mostly \equ{ds}. The reason is that there are more charged pions produced in the direct production, which compensates for the lower cross sections. Therefore, if \equ{ds} is used to estimate the neutrino flux from the photon flux coming from the $\pi^0$ decays, the neutrino fluxes are typically underestimated. From the center of mass energy dependence of the interactions, it can be shown that, independent of the input spectra, the ratio between charged pions and neutral pions is not 1:2, as in \equ{ds}, but at least 1.2:1~\cite{Hummer:2010vx}.

The neutrino production finally is given by the weak decays of pions and muons, as described in \Ref~\cite{Lipari:2007su}. Very importantly, the muon decays are helicity dependent, which means that the spin state of the muon has to be taken into account, and the four muon species $\mu^+_L$, $\mu^+_R$, $\mu^-_L$, and $\mu^-_R$ have to be treated separately.

\subsection{Neutrino fluxes}

For the observed neutrino fluxes, one typically distinguished three different conceptual cases. A \hl{single source} or \hl{point source} flux is related to a particular source, such as a GRB or AGN. Unfortunately, the statistics to be expected from a single (extragalactic) event is rather moderate, a few detected neutrinos at most; see, \eg, \Ref~\cite{Guetta:2003wi}. Therefore, different techniques have been proposed in the literature to increase the statistics.

The first approach is a \hl{stacking analysis}; see, \eg, \Ref~\cite{Becker:2005ej} for an example. In this case, the expected event rates from different, but similar sources are added. For example, the information from other messengers, such as gamma rays observed by BATSE or Fermi-LAT, can be used to compute the predicted neutrino flux under certain assumptions. By stacking the very few expected neutrino events from each source, reasonable statistics might be obtained with a good signal to background ratio, because time and directional information can be used to reduce the atmospheric neutrino background. One problem of the approach in \Ref~\cite{Becker:2005ej} is that the redshift of each source is needed to reconstruct the neutrino spectrum from an observed photon spectrum, which is only measured in a few cases. In addition, a few transient events seem to dominate the obtained neutrino prediction, which means that there may be selection effects.

The second possibility to increase statistics is measuring the \hl{diffuse flux} from all sources in the sky. A generic formula for the observed flux is given by~\cite{Becker:2007sv}
\begin{eqnarray}
\frac{dN_\nu}{dE_\nu}(E_\nu^0) & = & \int\limits_z \int\limits_{\mathcal{L}} \frac{d \Phi_\nu}{dE_\nu}(E_\nu^0, \mathcal{L}, z) \, \frac{d n}{d \mathcal{L} \, dz } (\mathcal{L}, z) \nonumber \\
& & \times  \frac{1}{4 \pi d_L(z)^2} \, . \label{equ:diffuse}
\end{eqnarray}
Here  $n$ is the source distribution function as a function of luminosity $\mathcal{L}$ and redshift $z$, $d \Phi_\nu/d E_\nu$ is the single source flux, and  $d_L$ is the luminosity distance.
The energy at the source $E_\nu$ is redshifted by the cosmic expansion to $E^0_\nu = E_\nu/(1+z)$.
Note that the source flux is already given in the observer's frame in this case. Diffuse fluxes provide the highest statistics, since all possible sources contribute. However, due to the lack of directional and timing information, the signal to background ratio may be poor. Especially at lower energies, the atmospheric neutrino background prohibits an extraction of the diffuse flux.

In summary, a point source flux provides the cleanest information, because there are no averaging effects involved. However, the statistics is poor. Stacked and diffuse fluxes increase the statistics significantly. It is however unclear what the impact of averaging is on the discussed measurements of neutrino properties.

\subsection{Flavor composition and propagation}

The discussion of particle physics properties of neutrinos often is based on the flavor composition at the source, which may be changed by propagation effects. Generically, three different source classes are distinguished, where neutrinos and antineutrinos are not discriminated:
\begin{description}
\item[Pion beam sources] produce neutrinos of the flavors $\nu_e$:$\nu_\mu$:$\nu_\tau$ in the flavor ratio 1:2:0, as it is expected from \equ{pion}.
\item[Muon damped sources] assume that the muons loose energy before they decay, which means that at high energies $\nu_e$:$\nu_\mu$:$\nu_\tau$ come in the flavor ratio 0:1:0 according to \equ{pion}.
\item[Neutron beam sources] assume neutrino production by neutron decays, which may come from the photo-dissociation of heavy nuclei. Therefore, one has $\nu_e$:$\nu_\mu$:$\nu_\tau$ in the flavor ratio 1:0:0.
\end{description}
Typically these flavor ratios are discussed in an energy-independent way. However, in practice, the flavor ratios change as a function of energy. For example, assume that neutrinos are produced by pion decays. In this case, the muons loose energy by synchrotron radiation already at lower energies than the pions because of the smaller mass. This means that the pion beam source changes into a muon damped source at the energy where the muon decay and cooling timescales are equal~\cite{Kashti:2005qa}, which is also sometimes called \hl{muon damping}. The actual energy dependence of the flavor ratios is non-trivial and depends not only on cooling and decay timescales, but also on the participating particle species and their interactions, see, \eg, \Refs~\cite{Kachelriess:2006fi,Lipari:2007su,Kachelriess:2007tr}. On the other hand, the energy dependence contains non-trivial information on the astrophysical properties.

The propagation of the neutrinos between source and detector consists of two parts. First of all, flavor mixing 
\begin{equation}
P_{\alpha \beta}= \sum\limits_{i=1}^3 | U_{\alpha i}|^2 |U_{\beta_i}|^2 \, ,
\label{equ:flmix}
\end{equation}
corresponding to neutrino oscillations with the oscillating part averaged out, changes the flavor ratios. The averaging of the neutrino oscillations can, for example, be justified by the size of the production region or decoherence on the way to the Earth. For example, if the original neutrinos are produced in the flavor ratio 
$\nu_e$:$\nu_\mu$:$\nu_\tau$ of 1:2:0, one will, depending on the mixing parameters, roughly have 1:1:1 at the Earth; see \Ref~\cite{Pakvasa:2008nx} and references therein. Note that \equ{flmix} is insensitive to CP violation, but it depends on the CP conserving part $\cos \deltacp$, which may be used to extract some information on the phase. The second part of propagation effects concerns with neutrinos passing the Earth's interior before being detected. At high energies, the Earth becomes opaque to $\nu_e$ and $\nu_\mu$ because of the interaction cross section increasing with energy. On the other hand, the $\nu_\tau$ are partly regenerated in the propagation.

\subsection{Neutrino detection and flavor ratios}

Neutrinos are detected with neutrino telescopes, which are primarily designed for the tracking of muons in water or ice coming from muon neutrino interactions. The muons and secondaries can be seen through Cherenkov radiation in photodetectors. The threshold for this process depends on the spacing of the photodetectors. As a peculiarity, the muons do not have to be produced in the actual detector volume or even pass it to be detected, only their light has to be received. The detection of muons is therefore described by an effective area depending to the actual detector geometry and the muon range, which is a function of energy.

Measuring flavor ratios of astrophysical neutrinos, on the other hand, is much more sophisticated (see \Ref~\cite{Beacom:2003nh} for an overview). Electron and tau neutrinos produce (apart from neutral current events) electrons and tauons, respectively, which result in showers (cascades) of particles. These can only be identified if the interaction vertex is within the detector volume. Therefore, these events are fiducial volume dominated, and the expected event rates are much lower. The discrimination between electromagnetic showers (from electrons) and hadronic showers (from tauons) is very difficult. At higher energies, however, the tauon may live long enough such that its track is separable, which means that the tau event may be identified as by so-called \hl{double bang} or \hl{lollipop} events~\cite{Learned:1994wg}, depending on the visibility of the first interaction vertex. In addition, the above mentioned Glashow resonance $\bar{\nu}_e + e^- \to W^- \to \hdots$ at about  $6.3 \, \text{PeV}$~\cite{Learned:1994wg,Anchordoqui:2004eb,Bhattacharjee:2005nh} may be used for (electron) antineutrino identification, whereas all other processes cannot distinguish between neutrinos and antineutrinos.

For a simplified (energy-independent) discussion, it turns out to be useful to define observables which fulfill the following two requirements:
\begin{enumerate}
\item
 They take into account the unknown flux normalization.
\item
 They take into account the detector properties.
\end{enumerate}
For requirement 1), \hl{flavor ratios},  \ie, the ratios of fluxes at the detector between different flavors, are the most popular choice, since in these ratios the flux normalization cancels. For requirement 2), it is convenient to define with increasing level of difficulty for the detection the following simplified ratios:
\begin{equation}
R = \frac{\phi_\mu^{\mathrm{Det}}}{\phi_e^{\mathrm{Det}}+\phi_\tau^{\mathrm{Det}}}  \, , 
\end{equation}
where $\phi_\alpha^{\mathrm{Det}}$ is the neutrino flux (accumulated over a certain energy range) of flavor $\nu_\alpha$ at the detector, neutrinos and antineutrinos summed over.
The ratio $R$ corresponds to the ratio of muon tracks to showers, where electromagnetic and hadronic showers do not need to be discriminated (here the background from neutral currents is ignored). Naturally, it is fiducial volume limited, \ie, the number of observed showers limits the statistics. As a second observable, we define
\begin{equation}
S = \frac{\phi_e^{\mathrm{Det}}}{\phi_\tau^{\mathrm{Det}}}  \, , 
\end{equation}
which basically corresponds to the number of electromagnetic (from $\nu_e$) to hadronic (from $\nu_\tau$) showers, or electromagnetic showers (from $\nu_e$) to double bang or lollipop events (from $\nu_\tau$) if the $\tau$ track can be resolved. Obviously, it is more difficult to measure $S$ than $R$, such as because of the limited statistics of the events which can be uniquely identified (for example double bang events). Furthermore, 
\begin{equation}
T = \frac{\hat \phi_e^{\mathrm{Det}}}{\phi_\mu^{\mathrm{Det}}}  
\end{equation}
represents the ratio of Glashow resonance electron antineutrino events (described by $\hat \phi_e^{\mathrm{Det}}$) to the muon tracks.
Note that these ratios do not take into account neutral current processes, different statistics between numerator and denominator, and backgrounds (such as from different $\tau$ decay modes), which means that results from these should be interpreted with care. 

These flavor ratios have been extensively used in the literature to identify a number of possible standard and new physics effects, see \Ref~\cite{Pakvasa:2008nx} for a review. Here we focus on two applications: Flavor mixing, which is a standard effect described by \equ{flmix}, and neutrino decay, which is too slow in the SM and therefore requires some new physics contribution.

\subsection{Constraints on neutrino flavor mixing}

The measurement of flavor mixings in \equ{flmix} has been extensively studied in the literature, see, \eg, \Refs~\cite{Bhattacharjee:2005nh,Farzan:2002ct,Beacom:2003zg,Serpico:2005sz,Serpico:2005bs,Winter:2006ce,Majumdar:2006px,Meloni:2006gv,Blum:2007ie,Awasthi:2007az,Rodejohann:2006qq,Xing:2006xd,Pakvasa:2007dc,Hwang:2007na,Choubey:2008di,Quigg:2008ab,Donini:2008xn,Xing:2008fg,Choubey:2009jq,Esmaili:2009dz,Lai:2010tj} in this context. Although the uncertainties in the parameters are probably too large for realistic applications, at least for pion beam sources~\cite{Meloni:2006gv}, there could be interesting information in special cases, such as using different sources or in combination with new physics scenarios, see, \eg, \Ref~\cite{Beacom:2003zg}.

Especially interesting may be the dependence of $R$, which is the flavor ratio possibly easiest to observe, on $\theta_{13}$ and $\deltacp$~\cite{Serpico:2005sz}. For example, one may expand $R$ for the
different astrophysical sources to first order in $\theta_{13}$~\cite{Winter:2006ce}
\begin{eqnarray}
R^{\mathrm{Neutron \, beam}} & \sim & 0.26 + 0.30 \, \,  \theta_{13} \, \cos \deltacp \, , \nonumber \\
R^{\mathrm{Muon \, damped}} & \sim & 0.66 - 0.52 \, \,  \theta_{13} \, \cos \deltacp  \, , \nonumber\\
R^{\mathrm{Pion \, beam}} & \sim & 0.50 - 0.14 \, \,  \theta_{13} \, \cos \deltacp \, . \nonumber \\  \label{equ:astro}
\end{eqnarray}
Higher order terms in $\theta_{13}$ are relatively small (but not negligible). 
Now compare this to superbeams, which are typically optimized for the the first oscillation
maximum. At the first oscillation maximum and in vacuum, one finds from \equ{papp}
\begin{equation}
P_{\mu e} \sim 2 \, \theta_{13}^2 \pm  0.09 \, \, \theta_{13} \, \sin \deltacp \, ,
\end{equation}
where the plus is for antineutrinos and the minus for neutrinos.
Obviously, these experiments are optimized for the observation of CP violation, which is proportional to $\sin \deltacp$ (the imaginary part of $\exp(i \deltacp)$), whereas the astrophysical sources can only capture the CP conserving part because of the averaging of the oscillations. Therefore, there is obviously complementary information.

\begin{figure}[t!]
\begin{center}
\includegraphics[width=\columnwidth]{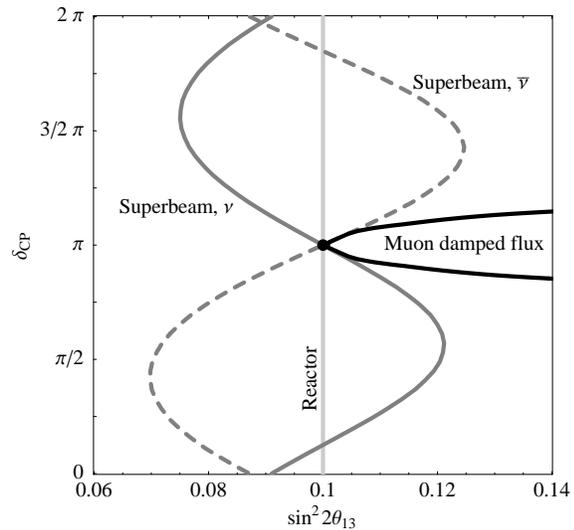}
\end{center}

\vspace*{-0.5cm}

\caption{\label{fig:didrates} Synergy among superbeams, reactor experiments, and astrophysical
fluxes (at the example of a muon damped source) in the $\stheta$-$\deltacp$-plane. Shown are the curves for constant total rates
(superbeam, reactor experiment) and constant R (astrophysical flux) going through the
best-fit point $\stheta=0.1$ and $\deltacp=\pi$. Figure taken from \Ref~\cite{Winter:2006ce}.
 }
\end{figure}

From \equ{astro}, we can read off that the modulation with $\deltacp$ is smallest for the pion beam source, which may be the most common case. Therefore, one can probably not expect any information from such a source. In the following, we will therefore focus on the other two cases. For example, consider \figu{didrates}, where the complementarity among superbeams, reactor experiments, and astrophysical sources is illustrated at the example of a muon damped source. In this case, the simulated rates and simulated $R$ are computed for $\stheta=0.1$ and $\deltacp=\pi$. The curves show the points in the $\stheta$-$\deltacp$-plane which produce exactly the same observables, \ie, are degenerate. Obviously, there is no information on $\deltacp$ from the reactor experiment alone, because the probabilities do not depend on $\deltacp$; \cf, \equ{rall}. Even if neutrinos and antineutrinos are used, there a degenerate solution remains for the superbeams. The astrophysical source alone, on the other hand, cannot limit $\stheta$, because there is a strong degeneracy with $\cos \deltacp$; \cf, \equ{astro}. The combination of superbeam or reactor experiment and astrophysical source, however, can in this case uniquely determine the point in the $\stheta$-$\deltacp$-plane. This synergy cannot only be used for the measurement of $\deltacp$, but also for the mass hierarchy, as illustrated in \Ref~\cite{Winter:2006ce}, because the degenerate solution in $\mathrm{sgn}(\ldm)$ moves in the $\deltacp$ direction as a function of $\stheta$~\cite{Huber:2002mx}. Note, however, that this complementarity is only relevant for large values of $\stheta$ and as long as there is no information on $\cos \deltacp$ from terrestrial sources, which can be obtained by wide band beams (beams with a broad energy spectrum), such as on-axis superbeams or  neutrino factories.

\begin{figure}[t!]
\begin{center}
\includegraphics[width=\columnwidth]{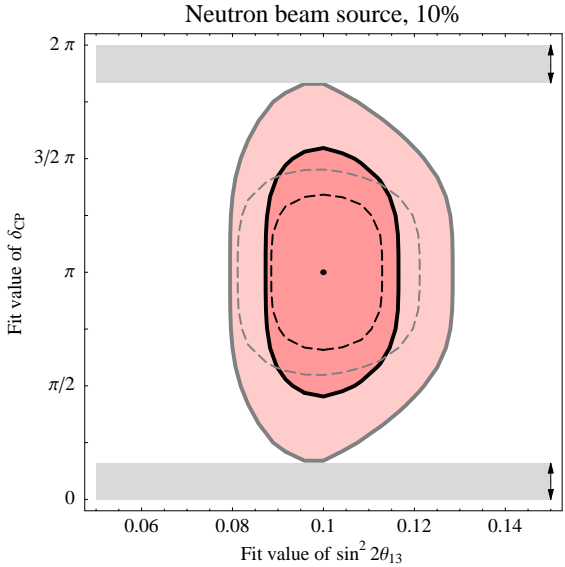}
\end{center}

\vspace*{-0.5cm}

\caption{\label{fig:reactorneu} Fit region 
 as function of $\stheta$ and $\deltacp$ for $R$, measured from a neutrino beam source at the level of 10\%, combined with Double Chooz. The simulated values
 are chosen as  marked by the dot. The contours are shown for the $1\sigma$ (black curves, dark regions) and 90\% (gray curves,
 light regions) confidence level (1 d.o.f.). Dashed curves represent the results when the other (not shown) oscillation
 parameters are fixed, \ie, not marginalized over. The arrows  mark the ranges in $\deltacp$ which can be
 excluded at the 90\% confidence level. Figure taken from \Ref~\cite{Winter:2006ce}.}
\end{figure}

The complementarity is shown at the example of a reactor experiment, sensitive to $\stheta$, and a neutron beam source, with $R$ measured at the level of 10\%, in \figu{reactorneu}. This example is, of course, somewhat optimistic, since a 10\% measurement of $R$ requires $\mathcal{O}(100)$ events already from the statistics point of view. However, it can be seen from this figure, that some information on $\deltacp$ is obtained.  Although there is no CP violation measurement in this case because the simulated value $\deltacp=\pi$  conserves the CP symmetry, a small range of $\deltacp$ around $0$ can be excluded at the 90\% confidence level.

\subsection{Constraints on neutrino decay}

\begin{figure*}[p!]
\begin{center}
 \includegraphics[width=0.65\textheight]{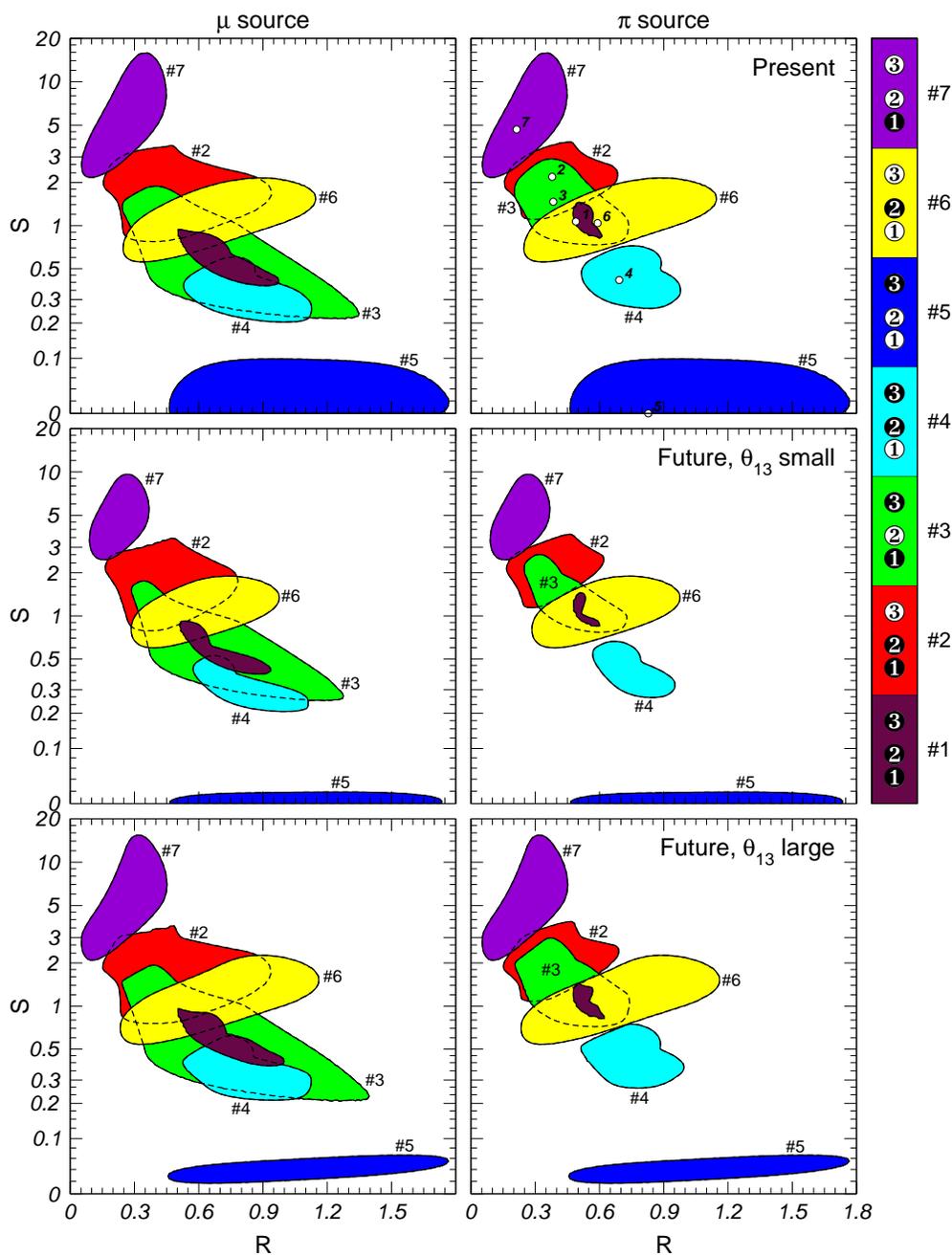}
\end{center}

\vspace*{-0.5cm}

\caption{\label{fig:flratios}     Allowed regions at 99\% CL in the $(R,\, S)$ plane corresponding
    to different complete decay scenarios, for a muon damped source (left
    panels) and a pion beam source (right panels). We assume a normal
    hierarchy. The upper panels correspond to the analysis of present
    data reported in \Ref~\cite{GonzalezGarcia:2007ib}. The other
    panels show the impact of 3 years of Double Chooz data taking (1.5
    with near detector), assuming no signal ($\stheta = 0$, middle
    panels) or a large signal ($\stheta=0.1$, lower panels). The 
    branching ratio parameters have been varied as well,
    where applicable. Figure taken from \Ref~\cite{Maltoni:2008jr}.}
\end{figure*}

Apart from ``conventional'' parameters, such as neutrino masses and
mixing angles, a very prominent example new physics properties
which can be tested by flavor ratios is the
neutrino lifetime; see, \eg, \Refs~\cite{Lipari:2007su,Beacom:2002vi,Maltoni:2008jr,Bhattacharya:2009tx}. Phenomenologically, from the observation of
neutrinos from supernova 1987A, we know that at least one neutrino
mass eigenstate must be stable over inter-galactic distances.
More stringent and explicit bounds can be derived from different
observations when specific decay models are assumed (see, \eg,
\Refs~\cite{Pakvasa:2008nx,Pakvasa:2003db,Yao:2006px} for an
overview). For example, solar neutrinos strongly limit the possibility
of radiative decays~\cite{Raffelt:1985rj}, while for Majoron
decays~\cite{Gelmini:1980re,Chikashige:1980qk} explicit bounds can be
obtained from neutrinoless double-beta decay and
supernovae~\cite{Tomas:2001dh}. Purely phenomenological (\ie,
model-independent) bounds are, however, much weaker, leaving enough
parameter space for the decay of any mass eigenstate over
extragalactic distances~\cite{Beacom:2002cb,Joshipura:2002fb,Bandyopadhyay:2002qg,GonzalezGarcia:2008ru}. 
Note that these bounds strongly depend on the mass eigenstate considered.
In view of the rather weak direct neutrino lifetime limits
and the recently proposed unparticle
models, which may lead to new mechanisms of neutrino
decay~\cite{Majumdar:2007mp,Chen:2007zy,Zhou:2007zq,Li:2007kj}, it may be well motivated 
to study the most general phenomenological case.

Conceptually, one may distinguish two types of neutrino decay~\cite{Lindner:2001fx}: \hl{Invisible decays} often refers to the decay into  invisible states, such as sterile neutrinos hardly mixing with the active ones, or unparticles. \hl{Visible decays} refers to decays in observable states, such as other active neutrinos. For example, in a degenerate mass scheme, invisible decays may be the most plausible assumption, because the decay kinematics depends on the mass differences. In a normal hierarchical spectrum, the heavier mass eigenstate $m_3$ may also decay into one of the two lighter ones. Another conceptual classification refers to the lifetime of the neutrinos. Depending on the lifetime and energy (see below), the neutrinos may have completely decayed over extragalactic distances, which is often called \hl{complete decays}. On the other hand, \hl{incomplete} decays show the characteristic energy dependence of decays.

Including incomplete invisible decays, the propagation effects in \equ{flmix} change into
\begin{equation}
P_{\alpha \beta}= \sum\limits_{i=1}^3 | U_{\alpha i}|^2 |U_{\beta_i}|^2 \, e^{- \alpha_i \frac{L}{E}} \, ,
\label{equ:flmixdecay}
\end{equation}
where $\alpha_i = m_i/\tau_i^0$ is the decay parameter as a function of the rest frame lifetime $\tau_i^0$. The characteristic energy dependence in the last term of \equ{flmixdecay} comes from the Lorentz boost of the rest frame lifetime $\tau_i = \tau_i^0 \, \gamma_i= \tau_i^0 \, E/m_i = E/\alpha_i$. 
For a recent discussion of incomplete decays as a function of energy in astrophysical fluxes, see \Ref~\cite{Bhattacharya:2009tx}. If the decay of mass eigenstate $j$ is complete, the exponential in the sum of \equ{flmixdecay} suppresses the $j$th term.
For a more refined discussion of possible combinations of neutrino oscillations and decay, including visible incomplete decays, see \Refs~\cite{Lindner:2001fx,Lindner:2001th}.

For complete decays, no matter if visible or invisible, it can be shown that only $2^3=8$ possibilities exist, because each of the three active mass eigenstates can be only stable or unstable. If all mass eigenstates are stable, we have the standard flavor mixing scenario in \equ{flmix}, if all mass eigenstates are unstable, no signal will be observed. Note that decay chains, such as $3 \rightarrow 2 \rightarrow 1$, can be integrated out in this approach, since eventually all $\nu_3$ and $\nu_2$ will end up as $\nu_1$. The eight possibilities discussed here are shown in the plot legend of \figu{flratios} on the right hand side. In this figure from \Ref~\cite{Maltoni:2008jr}, a filled disk represents a stable mass eigenstate, and an unfilled disk an unstable mass eigenstate. The different regions in the panels show the allowed regions at 99\% CL in the $(R,\, S)$ plane corresponding to the different complete decay scenarios in the legend, for a muon damped source (left
    panels) and a pion beam source (right panels). A normal
    hierarchy is assumed here.\footnote{In fact, there is some dependence on the mass hierarchy in such figures. Although the flavor mixing is insensitive to the mass hierarchy, the possible decay channels depend on the mass hierarchy. Therefore, the inverted hierarchy scenarios look somewhat different in general. } The upper panels correspond to the analysis of present
    data reported in \Ref~\cite{GonzalezGarcia:2007ib}. The other
    panels show the impact of 3 years of Double Chooz data taking (1.5
    with near detector), assuming no signal ($\stheta = 0$, middle
    panels) or a large signal ($\stheta=0.1$, lower panels). The 
    branching ratio parameters have been varied as well,
    where applicable. 

A particular measurement of $R$ and $S$ will lead to one specific point in the planes of \figu{flratios}. It is clear that some scenarios could not be disentangled easily, even if both $R$ and $S$ were measured to high precisions. However, if, for instance, a large value of $R$ is measured, then scenario \#5 can be easily identified even without measurement of $S$. In fact, it turns out that this observation is even robust for arbitrary sources without $\nu_\tau$ produced at the source. Therefore, the result at the end depends on the scenario actually implemented by nature. Furthermore, a future measurement of $\theta_{13}$ will somewhat reduce the regions, especially if $\theta_{13}$ is found to be small, as it can be read off from the middle row of the figure. Finally, it turns out that for this particular application the pion beam source (right column) is better suited, because it leads to smaller regions with respect to the oscillation parameter ranges. The reason is exactly the same as in the flavor mixing case: According to \equ{astro}, the pion beam is least affected by the unknown oscillation parameters (which is good for this application, but not so good for the flavor mixing measurement). 

\section{Determination of the absolute neutrino mass scale and the nature of neutrinos}
\label{sec:mass}

From \figu{spectrum}, we can read off that the spectrum of the neutrinos is fixed for known  mass squared differences $\sdm$ and $\ldm$, apart from an overall scale, the \hl{absolute neutrino mass scale}. This scale is determined by any of the three neutrino masses, such as the lightest neutrino mass $m_1$ (normal ordering -- NO) or $m_3$ (inverted ordering -- IO). In practice, it turns out that specific processes or classes of experiments are sensitive to particular combinations of the neutrino masses including the mixing angles and, for the case of $0\nu\beta\beta$ decay, also the Majorana phases. We emphasize the relationship between the observables and the absolute neutrino mass scale in this section. For more details, see, \eg, \Ref~\cite{Giunti:2007ry}.

The most direct access to the absolute neutrino mass is the test of the kinematical effect in nuclear beta decay, typically of tritium:
\begin{equation}
^3 \mathrm{H} \rightarrow ^3 \mathrm{He} + e^- + \bar \nu_e \, .
\label{equ:tritium}
\end{equation}
In this case, the endpoint of the electron spectrum depends on the mass of the neutrinos, no matter if Dirac or Majorana particles, but the effect is tiny.
The combination of neutrino masses, this experiment class is sensitive to, is given by
\begin{equation}
m_\beta^2 = \sum\limits_{i=1}^{3} | U_{e i} |^2 m_i^2 \, ,
\label{equ:mtrit}
\end{equation}
which comes from the incoherent sum over the contributions from different mass eigenstates in \equ{tritium}.
In terms of the neutrino oscillation observables and the lightest neutrino mass $m_1$ (normal ordering) or $m_3$ (inverted ordering), \equ{mtrit} can be expressed as
\begin{eqnarray}
m_\beta^2 & =  & m_1^2 + \Delta m_{21}^2 s_{12}^2 c_{13}^2 + \Delta m_{31}^2 s_{13}^2 \, \,  \mathrm{(NO)} \nonumber \\
m_\beta^2 & =  & m_3^2 + |\Delta m_{31}^2| c_{13}^2 + \Delta m_{21}^2 s_{12}^2 c_{13}^2 \, \,  \mathrm{(IO)} \nonumber \\
\end{eqnarray}
where $s_{ij} = \sin \theta_{ij}$ and $c_{ij} = \cos \theta_{ij}$. In degenerate case, the mass squared splittings are negligible, and the mixing effects can be neglected. In the normal hierarchical case ($m_1 \simeq 0$), the relative contribution of $m_3$ depends on $\theta_{13}$. For $\stheta \gtrsim 0.04$, the $m_3$ contribution dominates, for $\stheta \lesssim 0.04$, the $m_2$ contribution. For the inverted hierarchical case  ($m_3 \simeq 0$), both $m_1$ and $m_2$ contribute almost equally, with the ratio given by $\tan^2 \theta_{12}$. Note, however, that the hierarchical cases will not be accessible by the next generation of experiments.
The most stringent bound on $m_\beta$ was obtained by the Mainz~\cite{Kraus:2004zw} and Troitsk~\cite{Lobashev:1999tp} experiments, about $2 \, \mathrm{eV}$ (95\% CL). This limit is expected to be exceeded by one order of magnitude by the KATRIN experiment~\cite{Osipowicz:2001sq}.

Another very important test of the absolute neutrino mass scale is the $0\nu\beta\beta$ decay. In this case, a nucleus, for which the single beta decay is disfavored because of the pairing interaction in the nucleus, decays (for example) via
\begin{equation}
N(A,Z) \rightarrow N(A,Z+2) + 2 \, e^- \label{equ:0nbb}
\end{equation}
without emitting neutrinos. This processes can be associated with the Majorana neutrino mass. Simply speaking, there it corresponds to two simultaneous beta decay processes for which the outgoing neutrino of the first process becomes the incoming neutrino of the second processes.\footnote{Strictly speaking, \equ{weinberg} leads to $0\nu\beta\beta$ decay of this kind. However, other effective operators may induce $0\nu\beta\beta$ decay.
It can be shown that whatever operator causes $0\nu\beta\beta$ decay, also leads to \equ{weinberg}, and therefore Majorana neutrino masses, at the loop level; see generic argument in \Ref~\cite{Schechter:1981bd}.} This connection requires that the neutrino be its own antiparticle, \ie, has Majorana character. Because $0 \nu \beta \beta$ decay does not work in the SM, in which \equ{weinberg} is not present, an observation would probably one of the most exciting discoveries in neutrino physics. Note that the signature of $0\nu\beta\beta$ decay is very different from $2\nu\beta\beta$ (two simultaneous beta decays with neutrino emission). There are only two particles emitted in $0\nu\beta\beta$, namely the electrons, leading to fixed sum of the electron energies. For $2\nu\beta\beta$, one expects a continuous distribution of the sum of the electron energies, because the neutrinos can carry away energy.
\begin{figure}[t!]
\begin{center}
 \includegraphics[width=\columnwidth]{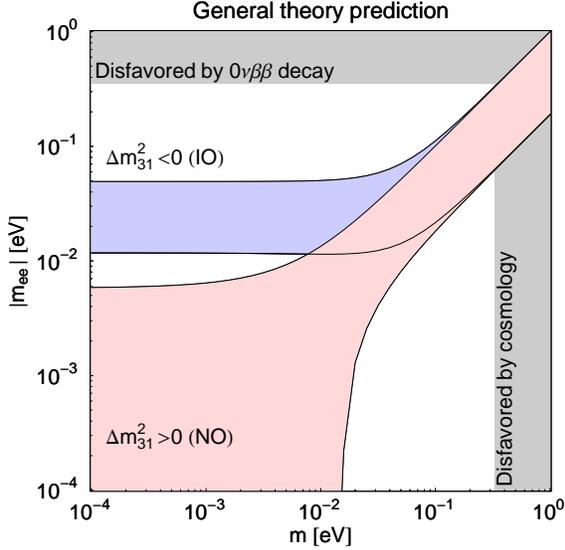}
\end{center}

\vspace*{-1cm}

\caption{\label{fig:0nbb}  
 General theory allowed region   for $|m_{ee}|$ in $0\nu\beta\beta$ decay as a function of the lightest neutrino mass $m$.We show the currently allowed regions for $|m_{ee}|$, where the mixing angles are varied in the current $3 \sigma$ ranges from \Tab~\ref{tab:summary} (computed with the formulas given in \Ref~\cite{Lindner:2005kr}).
Note that fixing $\theta_{13}$ would result in the appearance of the ``chimney''. 
The limit from cosmology is assumed do be $1 \, \mathrm{eV}$ for the sum of the three neutrino masses, and the the limit from $0\nu\beta\beta$-decay is obtained by the Heidelberg-Moscow collaboration~\cite{KlapdorKleingrothaus:2000sn}. 
We fix the mass squared differences to their best-fit values in \Tab~\ref{tab:summary}. Figure updated from \Ref~\cite{Plentinger:2006nb}.
}
\end{figure}
The combination of neutrino masses measured in $0 \nu \beta \beta$ decay is $\left| m_{ee} \right|^2$. With the definition of the Majorana phases in \equ{maj} it is given by (see, \eg, \Ref~\cite{Aalseth:2004hb} and references therein)
\be
\left| m_{ee} \right| \equiv \left| \sum U_{e i}^2 \, m_i \right| 
\ee
with
\be
|m_{ee}| = |m_{ee}^{(1)}| + |m_{ee}^{(2)}| \, e^{2 i \alpha} + 
|m_{ee}^{(3)}| \, e^{2 i (\beta-\deltacp)} 
\label{equ:mee}
\ee
and
\begin{eqnarray}
|m_{ee}^{(1)}| &=& m_1 \, |U_{e1}|^{2} = m_1 \, c_{12}^{2} \, c_{13}^{2} ~,
\nonumber\\
|m_{ee}^{(2)}| &=& m_2 \, |U_{e2}|^{2} = m_2 \, s_{12}^{2} \, c_{13}^{2} ~, \\
|m_{ee}^{(3)}| &=& m_3 \, |U_{e3}|^{2} = m_3 \, s_{13}^{2}~.\nonumber
\end{eqnarray}
In terms of the neutrino oscillation observables and the lightest neutrino mass $m_1$ (normal ordering) or $m_3$ (inverted ordering), \equ{mtrit} can be expressed as
\begin{eqnarray}
|m_{ee}| & = & m_1 c_{12}^2 c_{13}^2 + \sqrt{m_1^2+\Delta m_{21}^2} \, s_{12}^2 c_{13}^2 e^{2 i \alpha} \nonumber \\
& & + \sqrt{m_1^2+ \ldm } \, s_{13}^2 e^{2 i (\beta - \deltacp)} \, \, \mathrm{(NO)} \nonumber \\
|m_{ee}| & = & \sqrt{m_3^2+| \ldm|} \, c_{12}^2 c_{13}^2  \nonumber \\ & & + \sqrt{m_3^2+\sdm
+ |\ldm|} \, s_{12}^2 c_{13}^2 e^{2 i \alpha}  \nonumber \\ & & +m_3 \, s_{13}^2 e^{2 i (\beta - \deltacp)} \, \, \mathrm{(IO)}
\label{equ:meedet}
\end{eqnarray}
Obviously, these combinations strongly depend on the mass squared differences and mixing parameters. In addition, there are two new independent phases $\alpha$ and $\beta$, the Majorana phases.
As a peculiarity, for the inverted ordering in \equ{meedet}, the first term dominates even for small values of $m_3$, which means that $|m_{ee}|$ cannot vanish. For the normal ordering in \equ{meedet}, all the three terms could be of comparable magnitude. In this case, the three terms in \equ{meedet} can be pictured as complex numbers to be added as vectors in the complex plane, where the angles between the vectors are given by $2 \alpha$ and $2 (\beta -\deltacp)$ (see, \eg, \Ref~\cite{Lindner:2005kr}). If the vectors are of similar length, the three vectors may add to zero, leading to vanishing $0 \nu \beta \beta$ decay. Therefore, an observation of $0 \nu \beta \beta$ decay is not guaranteed, even if the neutrinos are Majorana particles. 
We illustrate the functional dependence of $|m_{ee}|$ in \equ{meedet} as a function of the lightest neutrino mass $m_1$ (NO) or $m_3$ (IO) in \figu{0nbb}. In this figure, the uncertainties of phases and mixing angles within their current $3 \sigma$ ranges are taken into account, which means that any point in the $0\nu\beta\beta$ regions could be allowed. For the NO, indeed $0\nu\beta\beta$ may be suppressed, whereas for the IO, a  non-vanishing  $|m_{ee}|$ is guaranteed. For large $m$, both cases are dominated by $m_1 \simeq m_3$, \ie, $|m_{ee}|$ increases linearly. Figures of this type are representative for presentations of $0\nu\beta\beta$ decay; however, the details depend on the parameter assumptions. In the future, especially the combination with the mass ordering measurements in long baseline experiments will be interesting. For example, the discovery of an inverted mass hierarchy at long baseline experiments in combination with improved $0 \nu \beta \beta$ bounds $|m_{ee}| \lesssim 0.01 \, \mathrm{eV}$ excluding the IO solution would point towards Dirac neutrino masses.
The current best bound $|m_{ee}| \lesssim 0.35 \, \mathrm{eV}$ (90\% CL) comes from the Heidelberg-Moscow collaboration~\cite{KlapdorKleingrothaus:2000sn}, where there is an uncertainty  coming from the calculations of the nuclear matrix elements (which we did not discuss here). An improvement of a factor of a few is expected, for example, from the GERDA experiment~\cite{Abt:2004yk}. The required sensitivity to exclude the IO solution might ultimatively be reached by the Majorana experiment~\cite{Aalseth:2004yt}.

Finally, indirect constraints to neutrino mass are nowadays obtained from cosmology. For example, too much mass in relativistic neutrinos prohibits the growth of small scale matter perturbations in the structure formation of the early universe, since energy is then transferred over longer scales. Cosmological tests of neutrino mass are typically sensitive to the sum of the neutrino masses $m = \sum_i m_i$. For example, the gravitational effect of the neutrino masses does not depend on their mixings. The current limit depends on the data sets included, and is of the order $1 \, \mathrm{eV}$; see \Ref~\cite{Hannestad:2010yi} for a very recent discussion. At this limit, neutrinos are quasi-degenerate, which means that $m_1 \simeq m_2 \simeq m_3 \simeq 1/3 \, \mathrm{eV}$, which is used in \figu{0nbb}.

\section{Summary and conclusions}

In these lectures, we have discussed the phenomenology of the most important neutrino properties, neutrino masses and mixings. In addition, we have shown a number of examples how to introduce new physics properties and how to test these in experiments. 

Most of the ``canonical'' neutrino properties are tested in neutrino oscillation experiments, namely the mixing angles, the Dirac CP phase, and the mass squared differences. We have shown that our current knowledge is based on two-flavor sub-sector measurements, often called the atmospheric oscillations, solar oscillations, and (short baseline) reactor oscillations. Of course, modern fits to data include the three-flavorness of neutrinos, but genuine three-flavor effects, such as leptonic CP violation, will be observable in the next generations of experiments. The key prerequisite for these measurements is the observation of $\theta_{13}$, see \Ref~\cite{Mezzetto:2010zi} for a recent review. Therefore, establishing $\theta_{13}>0$ is the current top priority in neutrino oscillation physics.  Beyond these canonical properties, future neutrino oscillation experiments could also probe new physics properties, such as coming from extra light sterile neutrinos or from LHC observable mediators. 

Apart from the mass squared differences, the absolute scale of the neutrino masses is needed to fix the mass spectrum. We have shown that different classes of experiments can probe the absolute neutrino mass scale, such as tritium endpoint experiments, $0\nu\beta\beta$ decay experiments, and cosmological tests. Whereas tritium endpoint experiments may be the most direct access to neutrino mass, cosmological tests, which we have only shortly touched, provide strong indirect constraints. Both of these tests work irrespective of the nature of neutrino mass, whereas the observation of $0\nu\beta\beta$ decay will point towards Majorana mass terms. Majorana mass terms may be especially interesting since they can be interpreted as lowest order perturbation of new physics. The simplest possibilities to obtain such Majorana mass terms are known as type~I,~II, or~III seesaw mechanisms, depending on the heavy mediator leading to the low energy mass term.

Except from neutrino sources such as the Earth's atmosphere, the Sun, or Earth-based experiments, we have also discussed cosmic accelerators, such as extragalactic AGNs and GRBs. Although the main interest in these sources may be of astrophysical origin, they can be used to probe particle physics properties of the neutrinos as well. In particular, the observation of flavor ratios in neutrino telescopes would help the extraction of such information. We have pointed out that in some extreme cases even information on neutrino flavor mixing may be obtained. It is, however, more likely that new physics properties, such as neutrino lifetime, will be probed using these sources. In order to illustrate the current state of research with respect to these measurements, he have discussed the sources, the propagation, and the detection, and we have sketched where current research requires further clarifications. For example, the properties of the source change the flavor ratios as a function of energy, which, on the one hand, obscures the cleanliness of the flavor ratio, but, on the other hand, implies additional information.

From the conceptual point of view, we have successively introduced neutrino oscillations in vacuum, neutrino oscillations in constant matter, and then neutrino oscillations in varying matter at the example of the Sun. Although we have only used the two-flavor limit, we have illustrated how the energy dependence of the solar oscillation effect can be reproduced in the simplest possible way. We have not included supernova neutrinos in the discussion, because collective effects make a simple analytical treatment difficult. For the neutrino properties, we have included masses and mixings and some new physics examples. For other neutrino properties, such as electromagnetic dipole moments, see also lectures by Z.-z.~Xing in this series~\cite{XingLecture}.

We conclude that the test of neutrino properties has two interesting components. First of all, the leptonic mixing angles, the Dirac CP phase and the mass squared differences will be tested by future experiments, which are neutrino properties expected in all currently accepted scenarios. In addition, the absolute neutrino mass scale needs to be determined or constrained, maybe even so strong that the inverted ordering can be excluded. The process of $0\nu\beta\beta$ decay  may also reveal the nature of neutrino mass, or constrain the contribution of Majorana masses for the inverted ordering. Apart from these guaranteed observations, neutrino physics may be one of the first places to search for new physics since massive neutrinos were not expected in the Standard Model. Especially future precision oscillation experiments, such as neutrino factories, and the observation of high energetic neutrino fluxes at neutrino telescopes may help to constrain such new physics properties, either because of precision (neutrino factories) or because of the high energies (neutrino telescopes). We therefore anticipate that neutrino physics could be good for further surprises.

\subsubsection*{Acknowledgments}

I would like to thank the organizers of the Schladming winter school 2010 for the invitation and for the excellent organization. In addition, I would like to thank Svenja H{\"u}mmer, Martin Krau{\ss}, Davide Meloni, and Jian Tang for comments on the manuscript.

%\bibliographystyle{h-elsevier}
%\bibliography{references}

\begin{thebibliography}{100}

\bibitem{Fukugita:1986hr}
M. Fukugita and T. Yanagida,
\newblock Phys. Lett. B174 (1986) 45.

\bibitem{Weinberg:1979sa}
S. Weinberg,
\newblock Phys. Rev. Lett. 43 (1979) 1566.

\bibitem{XingLecture}
Z.z. Xing,
\newblock Neutrino masses and flavor mixing,
\newblock Lecture given at the Schladming winter school 2010.

\bibitem{FritschLecture}
H. Fritsch,
\newblock Natural constants and their possible time variation,
\newblock Lecture given at the Schladming winter school 2010.

\bibitem{Cleveland:1998nv}
B.T. Cleveland et~al.,
\newblock Astrophys. J. 496 (1998) 505.

\bibitem{Bahcall:2004pz}
J.N. Bahcall, A.M. Serenelli and S. Basu,
\newblock Astrophys. J. 621 (2005) L85, astro-ph/0412440.

\bibitem{Bahcall:1976zz}
J.N. Bahcall and R. Davis,
\newblock Science 191 (1976) 264.

\bibitem{Ahmad:2002jz}
SNO, Q.R. Ahmad et~al.,
\newblock Phys. Rev. Lett. 89 (2002) 011301, nucl-ex/0204008.

\bibitem{Fukuda:1998mi}
Super-Kamiokande, Y. Fukuda et~al.,
\newblock Phys. Rev. Lett. 81 (1998) 1562, hep-ex/9807003.

\bibitem{GonzalezGarcia:2007ib}
M.C. Gonzalez-Garcia and M. Maltoni,
\newblock Phys. Rept. 460 (2008) 1, 0704.1800.

\bibitem{Giunti:2007ry}
C. Giunti and C.W. Kim,
\newblock Oxford, UK: Univ. Pr. (2007) 710 p.

\bibitem{Jenkins:2007ip}
E.E. Jenkins and A.V. Manohar,
\newblock Nucl. Phys. B792 (2008) 187, 0706.4313.

\bibitem{Giunti:1997wq}
C. Giunti and C.W. Kim,
\newblock Phys. Rev. D58 (1998) 017301, hep-ph/9711363.

\bibitem{Jarlskog:1985ht}
C. Jarlskog,
\newblock Phys. Rev. Lett. 55 (1985) 1039.

\bibitem{Albright:2006cw}
C.H. Albright and M.C. Chen,
\newblock Phys. Rev. D74 (2006) 113006, hep-ph/0608137.

\bibitem{PDG}
Particle Data Group, C. Amsler et~al.,
\newblock Phys. Lett. B667 (2008) 1.

\bibitem{Schwetz:2008er}
T. Schwetz, M.A. Tortola and J.W.F. Valle,
\newblock New J. Phys. 10 (2008) 113011, 0808.2016.

\bibitem{deGouvea:2008nm}
A. de~Gouvea and J. Jenkins,
\newblock Phys. Rev. D78 (2008) 053003, 0804.3627.

\bibitem{Fogli:2008jx}
G.L. Fogli et~al.,
\newblock Phys. Rev. Lett. 101 (2008) 141801, 0806.2649.

\bibitem{Maltoni:2008ka}
M. Maltoni and T. Schwetz,
\newblock PoS IDM2008 (2008) 072, 0812.3161.

\bibitem{GonzalezGarcia:2010er}
M.C. Gonzalez-Garcia, M. Maltoni and J. Salvado,
\newblock JHEP 04 (2010) 056, 1001.4524.

\bibitem{Harrison:2002er}
P.F. Harrison, D.H. Perkins and W.G. Scott,
\newblock Phys. Lett. B530 (2002) 167, hep-ph/0202074.

\bibitem{Aguilar:2001ty}
LSND, A. Aguilar et~al.,
\newblock Phys. Rev. D64 (2001) 112007, hep-ex/0104049.

\bibitem{Maltoni:2007zf}
M. Maltoni and T. Schwetz,
\newblock Phys. Rev. D76 (2007) 093005, 0705.0107.

\bibitem{Araki:2004mb}
KamLAND, T. Araki et~al.,
\newblock Phys. Rev. Lett. 94 (2005) 081801, hep-ex/0406035.

\bibitem{Apollonio:1999ae}
CHOOZ, M. Apollonio et~al.,
\newblock Phys. Lett. B466 (1999) 415, hep-ex/9907037.

\bibitem{Michael:2006rx}
MINOS, D.G. Michael et~al.,
\newblock Phys. Rev. Lett. 97 (2006) 191801, hep-ex/0607088.

\bibitem{Duchesneau:2002yq}
{OPERA}, D. Duchesneau,
\newblock eConf C0209101 (2002) TH09, hep-ex/0209082.

\bibitem{Mikheev:1985gs}
S.P. Mikheev and A.Y. Smirnov,
\newblock Sov. J. Nucl. Phys. 42 (1985) 913.

\bibitem{Mikheev:1986wj}
S.P. Mikheev and A.Y. Smirnov,
\newblock Nuovo Cim. C9 (1986) 17.

\bibitem{Wolfenstein:1978ue}
L. Wolfenstein,
\newblock Phys. Rev. D17 (1978) 2369.

\bibitem{Winter:2006vg}
W. Winter,
\newblock Earth Moon Planets 99 (2006) 285, physics/0602049.

\bibitem{Cervera:2000kp}
A. Cervera et~al.,
\newblock Nucl. Phys. B579 (2000) 17, hep-ph/0002108.

\bibitem{Freund:2001pn}
M. Freund,
\newblock Phys. Rev. D64 (2001) 053003, hep-ph/0103300.

\bibitem{Akhmedov:2004ny}
E.K. Akhmedov et~al.,
\newblock JHEP 04 (2004) 078, hep-ph/0402175.

\bibitem{Burguet-Castell:2001ez}
J. Burguet-Castell et~al.,
\newblock Nucl. Phys. B608 (2001) 301, hep-ph/0103258.

\bibitem{Minakata:2001qm}
H. Minakata and H. Nunokawa,
\newblock JHEP 10 (2001) 001, hep-ph/0108085.

\bibitem{Fogli:1996pv}
G.L. Fogli and E. Lisi,
\newblock Phys. Rev. D54 (1996) 3667, hep-ph/9604415.

\bibitem{Barger:2001yr}
V. Barger, D. Marfatia and K. Whisnant,
\newblock Phys. Rev. D65 (2002) 073023, hep-ph/0112119.

\bibitem{Huber:2003ak}
P. Huber and W. Winter,
\newblock Phys. Rev. D68 (2003) 037301, hep-ph/0301257.

\bibitem{ids}
International design study of the neutrino factory,
\newblock {\tt http://www.ids-nf.org}.

\bibitem{Bandyopadhyay:2007kx}
ISS Physics Working Group, A. Bandyopadhyay et~al.,
\newblock Rept. Prog. Phys. 72 (2009) 106201, 0710.4947.

\bibitem{Freund:1999gy}
M. Freund et~al.,
\newblock Nucl. Phys. B578 (2000) 27, hep-ph/9912457.

\bibitem{Ardellier:2006mn}
Double Chooz, F. Ardellier et~al.,
\newblock (2006), hep-ex/0606025.

\bibitem{Guo:2007ug}
Daya Bay, X. Guo et~al.,
\newblock (2007), hep-ex/0701029.

\bibitem{offaxis}
D. Beavis et~al.,
\newblock (1995).

\bibitem{Itow:2001ee}
Y. Itow et~al.,
\newblock Nucl. Phys. Proc. Suppl. 111 (2001) 146, hep-ex/0106019.

\bibitem{Ayres:2004js}
NOvA, D.S. Ayres et~al.,
\newblock (2004), hep-ex/0503053.

\bibitem{Ishitsuka:2005qi}
M. Ishitsuka et~al.,
\newblock Phys. Rev. D72 (2005) 033003, hep-ph/0504026.

\bibitem{Barger:2007yw}
V. Barger et~al.,
\newblock (2007), 0705.4396.

\bibitem{Zucchelli:2002sa}
P. Zucchelli,
\newblock Phys. Lett. B532 (2002) 166.

\bibitem{EURISOL}
Eurisol design study, beta beam task group,
\newblock {\tt http://cern.ch/beta-beam}.

\bibitem{Geer:1998iz}
S. Geer,
\newblock Phys. Rev. D57 (1998) 6989, hep-ph/9712290.

\bibitem{DeRujula:1998hd}
A. De~Rujula, M.B. Gavela and P. Hernandez,
\newblock Nucl. Phys. B547 (1999) 21, hep-ph/9811390.

\bibitem{Barger:1999fs}
V. Barger, S. Geer and K. Whisnant,
\newblock Phys. Rev. D61 (2000) 053004, \hfill \\ hep-ph/9906487.

\bibitem{Huber:2009cw}
P. Huber et~al.,
\newblock JHEP 11 (2009) 044, 0907.1896.

\bibitem{Huber:2007ji}
P. Huber et~al.,
\newblock Comput. Phys. Commun. 177 (2007) 432, hep-ph/0701187.

\bibitem{Huber:2004ka}
P. Huber, M. Lindner and W. Winter,
\newblock Comput. Phys. Commun. 167 (2005) 195, hep-ph/0407333,
\newblock {\tt http://www.mpi-hd.mpg.de/lin/globes/}.

\bibitem{Kopp:2006wp}
J. Kopp,
\newblock Int. J. Mod. Phys. C19 (2008) 523, physics/0610206.

\bibitem{Blennow:2009pk}
M. Blennow and E. Fernandez-Martinez,
\newblock Comput. Phys. Commun. 181 (2010) 227, 0903.3985.

\bibitem{Wilczek:1979hc}
F. Wilczek and A. Zee,
\newblock Phys. Rev. Lett. 43 (1979) 1571.

\bibitem{Buchmuller:1985jz}
W. Buchmuller and D. Wyler,
\newblock Nucl. Phys. B268 (1986) 621.

\bibitem{Grossman:1995wx}
Y. Grossman,
\newblock Phys. Lett. B359 (1995) 141, hep-ph/9507344.

\bibitem{Gonzalez-Garcia:2001mp}
M.C. Gonzalez-Garcia et~al.,
\newblock Phys. Rev. D64 (2001) 096006, hep-ph/0105159.

\bibitem{Bilenky:1992wv}
S.M. Bilenky and C. Giunti,
\newblock Phys. Lett. B300 (1993) 137, hep-ph/9211269.

\bibitem{Bergmann:1999pk}
S. Bergmann, Y. Grossman and D.M. Pierce,
\newblock Phys. Rev. D61 (2000) 053005, hep-ph/9909390.

\bibitem{Bilenky:1993bt}
M.S. Bilenky and A. Santamaria,
\newblock Nucl. Phys. B420 (1994) 47, hep-ph/9310302.

\bibitem{Cuypers:1996ia}
F. Cuypers and S. Davidson,
\newblock Eur. Phys. J. C2 (1998) 503, hep-ph/9609487.

\bibitem{Antusch:2008tz}
S. Antusch, J.P. Baumann and E. Fernandez-Martinez,
\newblock Nucl. Phys. B810 (2009) 369, arXiv:0807.1003.

\bibitem{Gavela:2008ra}
M.B. Gavela et~al.,
\newblock Phys. Rev. D79 (2009) 013007, arXiv:0809.3451.

\bibitem{Biggio:2009nt}
C. Biggio, M. Blennow and E. Fernandez-Martinez,
\newblock JHEP 08 (2009) 090, arXiv:0907.0097.

\bibitem{Tang:2009na}
J. Tang and W. Winter,
\newblock Phys. Rev. D80 (2009) 053001, 0903.3039.

\bibitem{Kopp:2008ds}
J. Kopp, T. Ota and W. Winter,
\newblock Phys. Rev. D78 (2008) 053007, 0804.2261.

\bibitem{Meloni:2009cg}
D. Meloni et~al.,
\newblock (2009), 0912.2735.

\bibitem{Abada:2007ux}
A. Abada et~al.,
\newblock JHEP 12 (2007) 061, 0707.4058.

\bibitem{Antusch:2006vwa}
S. Antusch et~al.,
\newblock JHEP 10 (2006) 084, hep-ph/0607020.

\bibitem{Antusch:2009pm}
S. Antusch et~al.,
\newblock Phys. Rev. D80 (2009) 033002, arXiv:0903.3986.

\bibitem{Giunti:2009en}
C. Giunti, M. Laveder and W. Winter,
\newblock Phys. Rev. D80 (2009) 073005, 0907.5487.

\bibitem{Altegoer:1997gv}
NOMAD, J. Altegoer et~al.,
\newblock Nucl. Instrum. Meth. A404 (1998) 96.

\bibitem{Eskut:1997ar}
CHORUS, E. Eskut et~al.,
\newblock Nucl. Instrum. Meth. A401 (1997) 7.

\bibitem{Raffelt:2007cb}
G.G. Raffelt and A.Y. Smirnov,
\newblock Phys. Rev. D76 (2007) 081301, 0705.1830.

\bibitem{Dasgupta:2007ws}
B. Dasgupta and A. Dighe,
\newblock Phys. Rev. D77 (2008) 113002, 0712.3798.

\bibitem{Lindner:2002wm}
M. Lindner et~al.,
\newblock Astropart. Phys. 19 (2003) 755, hep-ph/0207238.

\bibitem{Dighe:2003jg}
A.S. Dighe, M.T. Keil and G.G. Raffelt,
\newblock JCAP 0306 (2003) 006, hep-ph/0304150.

\bibitem{Alimonti:2000xc}
Borexino, G. Alimonti et~al.,
\newblock Astropart. Phys. 16 (2002) 205, hep-ex/0012030.

\bibitem{Learned:2000sw}
J.G. Learned and K. Mannheim,
\newblock Ann. Rev. Nucl. Part. Sci. 50 (2000) 679.

\bibitem{Becker:2007sv}
J.K. Becker,
\newblock Phys. Rept. 458 (2008) 173, 0710.1557.

\bibitem{Chiarusi:2009ng}
T. Chiarusi and M. Spurio,
\newblock Eur. Phys. J. C65 (2010) 649, 0906.2634.

\bibitem{Aslanides:1999vq}
ANTARES, E. Aslanides et~al.,
\newblock (1999), astro-ph/9907432.

\bibitem{Ahrens:2002dv}
IceCube, J. Ahrens et~al.,
\newblock Nucl. Phys. Proc. Suppl. 118 (2003) 388, astro-ph/0209556.

\bibitem{Mannheim:1993}
K. {Mannheim},
\newblock Astron. Astrophys. 269 (1993) 67, arXiv:astro-ph/9302006.

\bibitem{Mucke:2000rn}
A. Mucke and R.J. Protheroe,
\newblock Astropart. Phys. 15 (2001) 121, astro-ph/0004052.

\bibitem{Aharonian:2002}
F.A. {Aharonian},
\newblock Mon. Not. R. Astron. Soc. 332 (2002) 215, arXiv:astro-ph/0106037.

\bibitem{Waxman:1997ti}
E. Waxman and J.N. Bahcall,
\newblock Phys. Rev. Lett. 78 (1997) 2292, astro-ph/9701231.

\bibitem{Rachen:1998fd}
J.P. Rachen and P. Meszaros,
\newblock Phys. Rev. D58 (1998) 123005, astro-ph/9802280.

\bibitem{Waxman:1998yy}
E. Waxman and J.N. Bahcall,
\newblock Phys. Rev. D59 (1999) 023002, hep-ph/9807282.

\bibitem{Mannheim:1998wp}
K. Mannheim, R.J. Protheroe and J.P. Rachen,
\newblock Phys. Rev. D63 (2001) 023003, astro-ph/9812398.

\bibitem{Mucke:1999yb}
A. Mucke et~al.,
\newblock Comput. Phys. Commun. 124 (2000) 290, astro-ph/9903478.

\bibitem{Amsler:2008zzb}
Particle Data Group, C. Amsler et~al.,
\newblock Phys. Lett. B667 (2008) 1.

\bibitem{Hummer:2010vx}
S. Hummer et~al.,
\newblock (2010), 1002.1310.

\bibitem{Reynoso:2008gs}
M.M. Reynoso and G.E. Romero,
\newblock Astron. Astrophys. 493 (2009) 1, 0811.1383.

\bibitem{Learned:1994wg}
J.G. Learned and S. Pakvasa,
\newblock Astropart. Phys. 3 (1995) 267, hep-ph/9405296.

\bibitem{Anchordoqui:2004eb}
L.A. Anchordoqui et~al.,
\newblock Phys. Lett. B621 (2005) 18, hep-ph/0410003.

\bibitem{Bhattacharjee:2005nh}
P. Bhattacharjee and N. Gupta,
\newblock (2005), hep-ph/0501191.

\bibitem{Rachen:1996ph}
J.P. Rachen,
\newblock Interaction Processes and Statistical Properties of the Propagation
  of Cosmic Rays in Photon Backgrounds,
\newblock PhD thesis, MPIfR Bonn, Germany, 1996.

\bibitem{Kelner:2008ke}
S.R. Kelner and F.A. Aharonian,
\newblock Phys. Rev. D78 (2008) 034013, 0803.0688.

\bibitem{Lipari:2007su}
P. Lipari, M. Lusignoli and D. Meloni,
\newblock Phys. Rev. D75 (2007) 123005, 0704.0718.

\bibitem{Guetta:2003wi}
D. Guetta et~al.,
\newblock Astropart. Phys. 20 (2004) 429, astro-ph/0302524.

\bibitem{Becker:2005ej}
J.K. Becker et~al.,
\newblock Astropart. Phys. 25 (2006) 118, astro-ph/0511785.

\bibitem{Kashti:2005qa}
T. Kashti and E. Waxman,
\newblock Phys. Rev. Lett. 95 (2005) 181101, astro-ph/0507599.

\bibitem{Kachelriess:2006fi}
M. Kachelriess and R. Tomas,
\newblock Phys. Rev. D74 (2006) 063009, astro-ph/0606406.

\bibitem{Kachelriess:2007tr}
M. Kachelriess, S. Ostapchenko and R. Tomas,
\newblock Phys. Rev. D77 (2008) 023007, 0708.3047.

\bibitem{Pakvasa:2008nx}
S. Pakvasa,
\newblock Mod. Phys. Lett. A23 (2008) 1313, 0803.1701.

\bibitem{Beacom:2003nh}
J.F. Beacom et~al.,
\newblock Phys. Rev. D68 (2003) 093005, hep-ph/0307025.

\bibitem{Farzan:2002ct}
Y. Farzan and A.Y. Smirnov,
\newblock Phys. Rev. D65 (2002) 113001, hep-ph/0201105.

\bibitem{Beacom:2003zg}
J.F. Beacom et~al.,
\newblock Phys. Rev. D69 (2004) 017303, hep-ph/0309267.

\bibitem{Serpico:2005sz}
P.D. Serpico and M. Kachelriess,
\newblock Phys. Rev. Lett. 94 (2005) 211102, hep-ph/0502088.

\bibitem{Serpico:2005bs}
P.D. Serpico,
\newblock Phys. Rev. D73 (2006) 047301, hep-ph/0511313.

\bibitem{Winter:2006ce}
W. Winter,
\newblock Phys. Rev. D74 (2006) 033015, hep-ph/0604191.

\bibitem{Majumdar:2006px}
D. Majumdar and A. Ghosal,
\newblock Phys. Rev. D75 (2007) 113004, hep-ph/0608334.

\bibitem{Meloni:2006gv}
D. Meloni and T. Ohlsson,
\newblock Phys. Rev. D75 (2007) 125017, hep-ph/0612279.

\bibitem{Blum:2007ie}
K. Blum, Y. Nir and E. Waxman,
\newblock (2007), arXiv:0706.2070 [hep-ph].

\bibitem{Awasthi:2007az}
R.L. Awasthi and S. Choubey,
\newblock Phys. Rev. D76 (2007) 113002, 0706.0399.

\bibitem{Rodejohann:2006qq}
W. Rodejohann,
\newblock JCAP 0701 (2007) 029, hep-ph/0612047.

\bibitem{Xing:2006xd}
Z.z. Xing,
\newblock Phys. Rev. D74 (2006) 013009, hep-ph/0605219.

\bibitem{Pakvasa:2007dc}
S. Pakvasa, W. Rodejohann and T.J. Weiler,
\newblock JHEP 02 (2008) 005, 0711.4517.

\bibitem{Hwang:2007na}
G.R. Hwang and K. Siyeon,
\newblock Phys. Rev. D78 (2008) 093008, 0711.3122.

\bibitem{Choubey:2008di}
S. Choubey, V. Niro and W. Rodejohann,
\newblock Phys. Rev. D77 (2008) 113006, 0803.0423.

\bibitem{Quigg:2008ab}
C. Quigg,
\newblock (2008), 0802.0013.

\bibitem{Donini:2008xn}
A. Donini and O. Yasuda,
\newblock (2008), 0806.3029.

\bibitem{Xing:2008fg}
Z.z. Xing and S. Zhou,
\newblock Phys. Lett. B666 (2008) 166, 0804.3512.

\bibitem{Choubey:2009jq}
S. Choubey and W. Rodejohann,
\newblock Phys. Rev. D80 (2009) 113006, 0909.1219.

\bibitem{Esmaili:2009dz}
A. Esmaili and Y. Farzan,
\newblock Nucl. Phys. B821 (2009) 197, 0905.0259.

\bibitem{Lai:2010tj}
K.C. Lai, G.L. Lin and T.C. Liu,
\newblock (2010), 1004.1583.

\bibitem{Huber:2002mx}
P. Huber, M. Lindner and W. Winter,
\newblock Nucl. Phys. B645 (2002) 3, hep-ph/0204352.

\bibitem{Maltoni:2008jr}
M. Maltoni and W. Winter,
\newblock JHEP 07 (2008) 064, 0803.2050.

\bibitem{Beacom:2002vi}
J.F. Beacom et~al.,
\newblock Phys. Rev. Lett. 90 (2003) 181301, hep-ph/0211305.

\bibitem{Bhattacharya:2009tx}
A. Bhattacharya et~al.,
\newblock (2009), 0910.4396.

\bibitem{Pakvasa:2003db}
S. Pakvasa,
\newblock (2003), hep-ph/0305317.

\bibitem{Yao:2006px}
Particle Data Group, W.M. Yao et~al.,
\newblock J. Phys. G33 (2006) 1.

\bibitem{Raffelt:1985rj}
G.G. Raffelt,
\newblock Phys. Rev. D31 (1985) 3002.

\bibitem{Gelmini:1980re}
G.B. Gelmini and M. Roncadelli,
\newblock Phys. Lett. B99 (1981) 411.

\bibitem{Chikashige:1980qk}
Y. Chikashige, R.N. Mohapatra and R.D. Peccei,
\newblock Phys. Rev. Lett. 45 (1980) 1926.

\bibitem{Tomas:2001dh}
R. Tomas, H. Pas and J.W.F. Valle,
\newblock Phys. Rev. D64 (2001) 095005, hep-ph/0103017.

\bibitem{Beacom:2002cb}
J.F. Beacom and N.F. Bell,
\newblock Phys. Rev. D65 (2002) 113009, hep-ph/0204111.

\bibitem{Joshipura:2002fb}
A.S. Joshipura, E. Masso and S. Mohanty,
\newblock Phys. Rev. D66 (2002) 113008, hep-ph/0203181.

\bibitem{Bandyopadhyay:2002qg}
A. Bandyopadhyay, S. Choubey and S. Goswami,
\newblock Phys. Lett. B555 (2003) 33, hep-ph/0204173.

\bibitem{GonzalezGarcia:2008ru}
M.C. Gonzalez-Garcia and M. Maltoni,
\newblock Phys. Lett. B663 (2008) 405, arXiv:0802.3699 [hep-ph].

\bibitem{Majumdar:2007mp}
D. Majumdar,
\newblock (2007), arXiv:0708.3485 [hep-ph].

\bibitem{Chen:2007zy}
S.L. Chen, X.G. He and H.C. Tsai,
\newblock JHEP 11 (2007) 010, 0707.0187.

\bibitem{Zhou:2007zq}
S. Zhou,
\newblock Phys. Lett. B659 (2008) 336, 0706.0302.

\bibitem{Li:2007kj}
X.Q. Li et~al.,
\newblock Eur. Phys. J. C56 (2008) 97, 0707.2285.

\bibitem{Lindner:2001fx}
M. Lindner, T. Ohlsson and W. Winter,
\newblock Nucl. Phys. B607 (2001) 326, hep-ph/0103170.

\bibitem{Lindner:2001th}
M. Lindner, T. Ohlsson and W. Winter,
\newblock Nucl. Phys. B622 (2002) 429, astro-ph/0105309.

\bibitem{Kraus:2004zw}
C. Kraus et~al.,
\newblock Eur. Phys. J. C40 (2005) 447, hep-ex/0412056.

\bibitem{Lobashev:1999tp}
V.M. Lobashev et~al.,
\newblock Phys. Lett. B460 (1999) 227.

\bibitem{Osipowicz:2001sq}
KATRIN, A. Osipowicz et~al.,
\newblock (2001), hep-ex/0109033.

\bibitem{Schechter:1981bd}
J. Schechter and J.W.F. Valle,
\newblock Phys. Rev. D25 (1982) 2951.

\bibitem{Lindner:2005kr}
M. Lindner, A. Merle and W. Rodejohann,
\newblock Phys. Rev. D73 (2006) 053005, hep-ph/0512143.

\bibitem{KlapdorKleingrothaus:2000sn}
H.V. Klapdor-Kleingrothaus et~al.,
\newblock Eur. Phys. J. A12 (2001) 147, hep-ph/0103062.

\bibitem{Plentinger:2006nb}
F. Plentinger, G. Seidl and W. Winter,
\newblock Nucl. Phys. B791 (2008) 60, hep-ph/0612169.

\bibitem{Aalseth:2004hb}
C. Aalseth et~al.,
\newblock (2004), hep-ph/0412300.

\bibitem{Abt:2004yk}
I. Abt et~al.,
\newblock (2004), hep-ex/0404039.

\bibitem{Aalseth:2004yt}
Majorana, C.E. Aalseth et~al.,
\newblock Phys. Atom. Nucl. 67 (2004) 2002, hep-ex/0405008.

\bibitem{Hannestad:2010yi}
S. Hannestad et~al.,
\newblock (2010), 1004.0695.

\bibitem{Mezzetto:2010zi}
M. Mezzetto and T. Schwetz,
\newblock (2010), 1003.5800.

\end{thebibliography}

\end{document}